\documentclass[apj,twocolumn]{emulateapj_mod}
\usepackage{epsfig,apjfonts,mathptmx}

\makeatother

\newcommand{\lsim}{\ \raise -2.truept\hbox{\rlap{\hbox{$\sim$}}\raise5.truept
        \hbox{$<$}\ }}
\newcommand{\gsim}{\ \raise -2.truept\hbox{\rlap{\hbox{$\sim$}}\raise5.truept
        \hbox{$>$}\ }} 
\newcommand{\kms}{km~s$^{-1}$}
\newcommand{\cm}{cm$^{-2}$}
\newcommand{\feii}{$\log N_{\rm FeII}$}
\newcommand{\mnii}{$\log N_{\rm MnII}$}
\newcommand{\mgii}{$\log N_{\rm MgII}$}
\newcommand{\mgi}{$\log N_{\rm MgI}$}

\newcommand{\siii}{$\log N_{\rm SiII}$}

\newcommand{\feiib}{$\log N_{\rm FeII}=15.54^{+0.23}_{-0.13}$}
\newcommand{\mniib}{$\log N_{\rm MnII}=13.49\pm0.13$}
\newcommand{\mgib}{$\log N_{\rm MgI}=12.93\pm0.13$}
\newcommand{\mgiib}{$\log N_{\rm MgII} = 15.26^{+0.33}_{-0.24}$}

\newcommand{\zi}{$1.3<z<2.4$}
\newcommand{\zii}{$1.34<z<1.97$}
\newcommand{\zav}{1.60} 
\newcommand{\zme}{1.56} 

\newcommand{\coadd}{composite}

\slugcomment{ApJ in press}
\shorttitle{GDDS: Metals in $1.3<z<2$ Galaxies}
\shortauthors{Savaglio et al.}

\begin{document} 

\input epsf 

\title{THE GEMINI DEEP DEEP SURVEY: II. 
METALS IN STAR--FORMING GALAXIES AT REDSHIFT $1.3<z<2$\altaffilmark{1}}

\author{S. Savaglio\altaffilmark{2,3}, 
K. Glazebrook\altaffilmark{2},
R. G. Abraham\altaffilmark{4},
D. Crampton\altaffilmark{5},
H.--W. Chen\altaffilmark{6,7},
P. J. P. McCarthy\altaffilmark{8},
I. J\o rgensen\altaffilmark{9}, 
K. C. Roth\altaffilmark{9},
I. M. Hook\altaffilmark{10},
R. O. Marzke\altaffilmark{11},
R. G. Murowinski\altaffilmark{5},
R. G. Carlberg\altaffilmark{4}
}

\altaffiltext{1}{Based on observations obtained with the Gemini North
Telescope}

\altaffiltext{2}{Johns Hopkins University, 3400 North Charles Street,
Baltimore, MD21218; savaglio@pha.jhu.edu}

\altaffiltext{3}{On leave of absence from Osservatorio Astronomico di
Roma, Italy} 

\altaffiltext{4}{Department of Astronomy \& Astrophysics, University
of Toronto, 60 St. George Street, Toronto, ON, M5S 3H8, Canada}

\altaffiltext{5}{Herzberg Institute of Astrophysics, National Research Council,
5071 West Saanich Road, Victoria, British Columbia, V9E 2E7, Canada}

\altaffiltext{6}{Center for Space Research, Massachusetts Institute of
Technology, Cambridge, MA 02139-4307}

\altaffiltext{7}{Hubble Fellow} 

\altaffiltext{8}{Observatories of the Carnegie Institution of Washington, 813
Santa Barbara Street, Pasadena, CA 91101}

\altaffiltext{9}{Gemini Observatory, Hilo, HI 96720}

\altaffiltext{10}{Department of Astrophysics, Nuclear \& Astrophysics
Laboratory, Oxford University, Keble Road, Oxford OX1 3RH, UK}

\altaffiltext{11}{Department of Physics and Astronomy, San Francisco State
University, 1600 Holloway Avenue, San Francisco, CA 94132}

\altaffiltext{12}{Herzberg Institute of Astrophysics, National
Research Council, 5071 West Saanich Road, Victoria, British Columbia,
V9E 2E7, Canada}

\altaffiltext{13}{Department of Astronomy \& Astrophysics, University
of Toronto, 60 St. George Street, Toronto, ON, M5S 3H8, Canada}

\begin{abstract} 

The goal of the Gemini Deep Deep Survey (GDDS) is to study an unbiased
sample of $K<20.6$ galaxies in the redshift range $0.8<z<2.0$.  Here
we determine the statistical properties of the heavy element
enrichment in the interstellar medium (ISM) of a subsample of 13
galaxies with \zii~and UV absolute magnitude $M_{2000}<-19.65$.  The
sample contains 38\% of the total number of identified galaxies in the
first two fields of the survey with $z>1.3$. The selected objects have
colors typical of irregular and Sbc galaxies.  Strong [\ion{O}{2}]
emission indicates high star formation activity in the \ion{H}{2}
regions (SFR $\sim13-106$ M$_\odot$ yr$^{-1}$). The high S/N composite
spectrum shows strong ISM \ion{Mg}{2} and \ion{Fe}{2} absorption,
together with weak \ion{Mn}{2} and \ion{Mg}{1} lines. The \ion{Fe}{2}
column density, derived using the curve of growth analysis, is \feiib.
This is considerably larger than typical values found in damped
Ly$\alpha$ systems (DLAs) along QSO sight lines, where only 10 out of
87 ($\sim11$\%) have \feii~$\geq 15.2$. High \ion{Fe}{2} column
densities are observed in the $z=2.72$ Lyman break galaxy cB58
(\feii~$\simeq 15.25$) and in gamma--ray burst host galaxies
(\feii~$\sim 14.8-15.9$).  Given our measured \ion{Fe}{2} column
density and assuming a moderate iron dust depletion ($\delta_{\rm
Fe}\sim 1$ dex), we derive an optical dust extinction $A_V\sim 0.6$.
If the \ion{H}{1} column density is $\log N_{\rm HI}<21.7$ (as in 98\%
of DLAs), then the mean metallicity is $Z/Z_\odot > 0.2$.  The high
completeness of the GDDS sample implies that these results are typical
of star--forming galaxies in the $1<z<2$ redshift range, an epoch
which has heretofore been particularly challenging for observational
programs.

\end{abstract}


\section{Introduction}

The interstellar medium (ISM) is an important component in galaxies,
not only in terms of mass, but also because it carries information on
past generations of stars, and provides the main fuel for the new
generations of stars. In the local Universe the cold ISM ($T\lsim
1000$ K) is $\sim$15-40\% of the total mass in stars (Fukugita et
al.~1998; Zwaan et al.~2003), and at higher redshifts it is a major
constituent of the Universe ($\sim 2$ times higher than in the local
Universe at $1.5<z<2.0$; Storrie--Lombardi \& Wolfe 2000; Peroux et
al.~2003).  In the cold ISM, heavy elements -- the footprint of the
stellar activity -- can most easily be investigated through the
analysis of UV absorption lines.

Apart from very detailed analysis in Galactic ISM (Savage \& Sembach
1996), in the local Universe exploring the UV absorption lines become
possible only for a few bright star-forming galaxies (Heckman et
al.~2001; Aloisi et al.~2003) due to limited capabilities of UV
satellites. At higher redshifts, bright QSO sources, whose sight lines
cross intervening galaxies, allow the detection of absorption lines
with UV satellites and (at $z>2.5$) ground-based telescopes. QSO
damped Lyman--$\alpha$ systems (QSO--DLAs) have probed the ISM in more
than 100 galaxies at $0.0<z<4.5$ (Prochaska et al.~2003a). These data
may be biased, however, to galaxies or regions of galaxies that are
dust- and metal-poor since dusty, metal rich systems are likely to
obscure or, at least, extinct the background QSO.

Indeed, we have indications that the ISM in other high redshift
objects is chemically more evolved.  For instance, gamma--ray burst
(GRB) afterglows are associated with regions of intense star formation
(Bloom, Kulkarni, \& Djorgovski 2002; Djorgovski et al. 2003). They are
-- for a short time -- much brighter than QSOs ($\sim 1000$ times
brighter in X--ray fluxes) and can be detected even if embedded in a
very dense medium. The analysis of UV absorption lines in 4 GRB
spectra (GRB--DLAs) shows high column density of metals in the ISM 
of the host galaxies (Savaglio, Fall, \& Fiore 2003; Vreeswijk et
al.~2003).  In addition, large surveys of $z>2.5$ Lyman--break
galaxies (LBGs, Steidel et al.~2003) show predominantly high
equivalent widths (EWs) of ISM metal absorption lines (Pettini et
al.~2002; Shapley et al.~2003). 

The direct study of cold ISM in galaxies requires reasonable S/N of
the rest--frame UV continuum, and this is generally not
sufficiently strong for our observational capabilities, even when
redshifted to the optical at high $z$.  Additionally, systematic
problems arise from imperfect sky subtraction and CCD fringe
removal. As a consequence, surveys have mainly targeted $z>2.5$
galaxies where the Lyman--break technique starts to be effective, or
strong emission line objects, observable in the optical up to
$z\sim1$. The $1<z<2.5$ range is often referred to as the ``redshift
desert'' for lack of discoveries. This interval is also very
problematic for studies of galaxy evolution, since it apparently spans
a major epoch of galaxy building: 50 -- 75\% of the stellar mass in
galaxies was in place by $z\sim1$, whereas at $2.5<z<3.5$ the Lyman
break galaxies contain only $\sim 3-20$\% of the present day stellar
mass (Dickinson et al.~2003; Fontana et al.~2003).  Recently, the
$K20$ survey (Cimatti et al.~2002a) with 82 spectroscopically
confirmed $K<20$ galaxies at $1 < z < 2$ have started to unveil
some aspects of galaxy evolution in the redshift desert (Cimatti et
al.~2002b; Pozzetti et al.~2003). However, the number of galaxies
drops to 17 if the $1.3 < z<2$ interval is considered.

To improve optical spectroscopy, a new observing mode known as Nod \&
Shuffle (N\&S), has been recently developed (Glazebrook \&
Bland-Hawthorn 2001; Cuillandre et al.~1994).  During N\&S
observations, the telescope is rapidly switched between object and sky
positions (``nodding''), while the charge is ``shuffled'' between
different parts of the CCD.  Because both the sky and objects are
observed through the same optical path, slits, and pixels, the effects
of fringing are eliminated and the sky subtraction greatly improved.
As a result, N\&S allows very long integration times before systematic
effects start to dominate, $\sim10$ times longer than is practical
with conventional spectroscopy.

We have begun a project to observe galaxies in the redshift desert
(the Gemini Deep Deep Survey, GDDS, Abraham et al.~in preparation)
that uses the N\&S technique on the multi--object spectrograph at the
Gemini 8m telescope.  The $\sim 10^5$ seconds exposures allow us to
reach very faint magnitudes (limiting magnitude $I_{AB}=24.8$ and
$K=20.6$, more than half a magnitude deeper than $K20$).  Both raw and
fully reduced data from the GDDS will be publicly
available\footnote{Data can be downloaded at the GDDS homepage
http://www.ociw.edu/lcirs/gdds.html}. In this paper we present first
results on the heavy element enrichment in the cold ISM of a subset of
galaxies at $1.3 < z < 2.0$, based on observations of the first two
fields.  We estimate the column densities of \ion{Fe}{2}, \ion{Mn}{2},
and \ion{Mg}{2} using the curve of growth analysis, and compare our
results with what is already known of the ISM over a larger redshift
scale.

The paper is structured as follows: in \S 2 we present the data set,
the sample selection is described in \S 3, in \S 4 we report
[\ion{O}{2}] line flux measurements, in \S 5 we describe the method
used to combine the individual spectra, in \S 6 we measure column densities
of  metals, in \S 7 we report results of other surveys, in \S 8 we
discuss the general properties of the selected sample, and in \S 9 we
summarize the main conclusions. Throughout the paper we adopt a $h
\equiv H_o/100= 0.7$, $\Omega_M = 0.3$, $\Omega_\Lambda = 0.7$
cosmology.

\begin{table*}
\footnotesize
\begin{center}
\caption{Selected galaxy sample}
\begin{tabular}{lcccccccccc}
\tableline\tableline&&&&&&&&&&\\[-5pt]
GDDS ID & $z_{\rm ISM}$\,\tablenotemark{a} & $z_{\rm [OII]}$\,\tablenotemark{b}
 & $\Delta v$\,\tablenotemark{c} & $K$ & $I-K$ & $M_{2000}$\,\tablenotemark{d}
& $F_{\rm [OII]}$\,\tablenotemark{e} & $L_{\rm [OII]}$ &
SFR$_{\rm [OII]}$\,\tablenotemark{f} & SFR$_{2000}$\,\tablenotemark{e} \\
& & & (\kms) &  & & & ($10^{-17}$ erg s$^{-1}$ \cm) & ($10^{41}$ erg s$^{-1}$) & (M$_\odot$ yr$^{-1}$) & (M$_\odot$ yr$^{-1}$) \\ 
[5pt]\tableline &&&&&&&&&&\\[-5pt] 
02-1417 & 1.5998 & 1.5990 & $-92$    & $19.85\pm0.35$ & $3.48\pm0.35$ & $-19.70$ & $3.19\pm0.73$  & $5.3\pm1.2$  & 13 & 20 \\
02-1636 & 1.6357 & 1.6360 & $+34$ & $20.24\pm0.42$ & $3.07\pm0.42$ & $-20.03$ & $9.74\pm0.84$  & $17.1\pm1.5$ & 43 & 27 \\
02-1790 & 1.5778 & 1.5767 & $-128$  & $20.64\pm0.52$ & $2.36\pm0.52$ & $-20.70$ & $26.00\pm0.70$ & $41.7\pm1.1$ & 106 & 49 \\
02-2530 & 1.5278 & 1.5263 & $-178$  & $20.32\pm0.44$ & $2.67\pm0.44$ & $-20.17$ & $9.40\pm0.69$  & $13.9\pm1.0$ & 35 & 30 \\
22-0964 & 1.5113 & 1.5124 & $+131$ & $19.61\pm0.28$ & $3.30\pm0.28$ & $-19.81$ & $4.61\pm0.58$  & $6.7\pm0.8$  & 17 & 22 \\
22-1042 & 1.5228 & 1.5248 & $+237$ & $20.00\pm0.34$ & $3.17\pm0.34$ & $-19.69$ & $7.99\pm0.51$  & $11.8\pm0.8$ & 30 & 20 \\ 
22-1055 & 1.3407 & 1.3410 & $+38$ & $19.47\pm0.26$ & $3.39\pm0.26$ & $-19.70$ & $4.21\pm0.15$  & $4.5\pm0.2$  & 12 & 20 \\ 
22-1559 & 1.8954 & . . .  & . . .  & $20.74\pm0.49$ & $2.97\pm0.49$ & $-20.56$ &  . . .  & . . .&  . . . & 44 \\ 
22-1909 & 1.4847 & 1.4876 & $+350$ & $20.81\pm0.52$ & $2.33\pm0.52$ & $-20.43$ & $8.62\pm0.31$  & $12.0\pm0.4$ & 30 & 39  \\
22-2172 & 1.5613 & 1.5613 & $0$  & $20.41\pm0.41$ & $2.69\pm0.41$ & $-20.29$ & $11.86\pm0.56$ & $18.6\pm0.9$ & 47 & 34 \\ 
22-2264 & 1.6703 & 1.6742 & $+438$ & $20.60\pm0.46$ & $2.82\pm0.47$ & $-20.35$ & $17.03\pm1.55$ & $31.6\pm2.9$ & 80 & 36 \\ 
22-2395 & 1.4847 & 1.4863 & $+193$ & $20.45\pm0.42$ & $2.46\pm0.42$ & $-20.80$ & $5.08\pm0.34$  & $7.0\pm0.5$  & 18 & 54 \\ 
22-2400 & 1.9669 & . . .  & . . .  & $20.34\pm0.41$ & $3.14\pm0.42$ & $-21.07$ & . . .  & . . . & . . . & 69 \\  
[2pt]\tableline\\
\end{tabular}
\vskip 7pt
\begin{minipage}{15cm}
\tablenotetext{a}{Redshift of the ISM absorption lines.}
\tablenotetext{b}{Redshift of the [O II] emission line.}
\tablenotetext{c}{Velocity difference between $z_{\rm [OII]}$ and $z_{\rm ISM}$.}
\tablenotetext{d}{AB absolute magnitude in a synthetic box filter centered at $\lambda=2000$ \AA. The accuracy is $\sim 0.2$ magnitudes (see text).}
\tablenotetext{e}{The [\ion{O}{2}] fluxes have been measured from the spectra, and no aperture correction has been applied. Errors include only photon counting statistics.} 
\tablenotetext{f}{SFRs are estimated assuming SFR$_{\rm [OII]}({\rm M_\odot~yr^{-1}}) \simeq 2.54\times L_{\rm [OII]} \times 10^{-41}$ erg s$^{-1}$ cm$^{-2}$.}
\tablenotetext{g}{SFRs are estimated assuming SFR$_{2000}({\rm M_\odot~yr^{-1}})\simeq 1.25 \times L_{2000} \times 10^{-28}$ erg s$^{-1}$ Hz$^{-1}$.}
\end{minipage}
\end{center}
\end{table*}

\section{GDDS data}

The spectra presented in this paper were obtained during three
observing runs at the Gemini North telescope in the period 2002
August--October. The two fields SA22 ($\alpha_{2000}= 22^h17^m20^s$,
$\delta_{2000}= +00^\circ21'30''$) and NDWS\footnote{NDWS: NOAO Deep
Wide-Field Survey.}  ($\alpha_{2000}= 02^h10^m00^s$,
$\delta_{2000}=-04^\circ30'00''$) were selected from the Las Campanas
Infrared (LCIR) Survey (McCarthy et al.~2001; Chen et al.~2002; Firth
et al.~2002; Chen et al.~2003), and observed with the Gemini
multi-object spectrograph (GMOS, $5.5'\times5.5'$ field of view;
Murowinski et al.~in preparation; Hook et al.~2003). The ``queue
scheduled'' mode at the Gemini observatory ensured that data were
taken only when the weather conditions were as requested. As a result,
during our observations the seeing was $\lsim 0.85''$ in $I$ band for
most of the time. The LCIR Survey photometry ($VIzHK$ filters for
field SA22 and $BVRIzK$ filters for field NDWS) was used to preselect
$z>0.8$ objects in the two fields with the photometric redshift
technique.  Two GMOS masks were used for field SA22 (exposure times
48600 and 90000 seconds) with 111 slits total, and one mask with 60
slits for field NDWS (exposure time 75600 seconds). The total number
of objects observed was then 171.  The observations were taken using
$0.75''\times2.2''$ and $0.75''\times2''$ slit apertures. The
comparison with similarly faint IR galaxies at $z>1$ (Yan et al. 1998)
suggests that the slit apertures allow to collect more than 50\% of
the galaxy flux.  The 2D data reduction was performed using
IRAF\footnote{IRAF is distributed by the National Optical Astronomy
Observatories, which are operated by the Association of Universities
for Research in Astronomy, Inc., under cooperative agreement with the
National Science Foundation.} and Gemini/GMOS tools\footnote{For more
information visit \\
http://www.gemini.edu/sciops/instruments/gmos/gmosIndex.html}.  For
the 1D spectrum extractions we used iGDDS, a package specifically
written with the toolkit Cocoa under Mac OS X (details in Abraham et
al.~in preparation). Uncertainty in the wavelength calibration is of
the order of 1.5 \AA.  The wavelength calibrated spectra were
corrected for the telluric absorption and flux calibrated. Most of the
observations were carried out near transit and no corrections for
differential atmospheric refraction (which is small at these
wavelengths) were applied.  To confirm that the N\&S technique was
working as expected, sky residuals were measured from some strong
lines in a stack of 60 spectra. The residuals are of the order of
$0.05-0.1$\%, as predicted by Glazebrook \& Bland-Hawthorn (2001).
The final spectral coverage in each spectrum is within the
$5000-10000$ \AA~interval, depending on the position of the slit in
the mask, and the dispersion is $\Delta\lambda\simeq3.5$ \AA~per
pixel. The resolution element, determined by the slit width, is 4.5
pixels or FWHM $\simeq 16$ \AA, equivalent to $960-460$ \kms, from the
blue to the red end of the spectral range. The resolution was also
measured from unresolved sky emission lines to be FWHM $\simeq 17-18$
\AA, slightly larger than the nominal resolution. This is not
unexpected since the sky emission is extended over the whole slit
aperture.

The spectra were extracted both using the linear and the optimal
extraction method.  In parallel the noise spectrum was created for
each object using the object and sky extracted signal, and the
read--out--noise of the instrument ($3-3.5~e^-$, depending on the
CCD). Redshifts were found from simple eye inspection of line
positions.  Manual redshifting (possible thanks to the small number of
spectra) was done because of the ability of the eye to reject
extraneous data artifact features in the 2D spectrum. Among the 171
objects observed, 7 have been identified as stars, 4 as AGNs, 18 have
tentative redshift, 18 have no redshift.  The remaining 124
extragalactic objects have secure redshift.

\begin{figure}\label{f0}
\centerline{{\epsfxsize=9cm \epsfbox{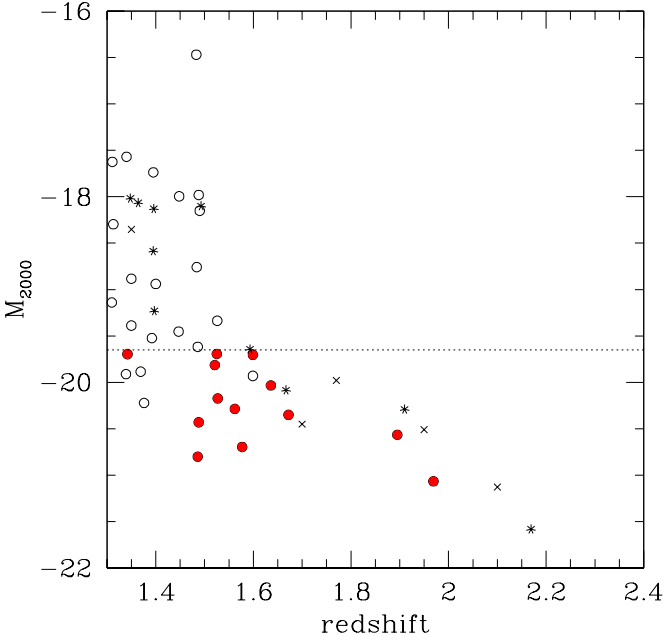}}}
\figcaption[f0]{AB $M_{2000}$ absolute magnitude of GDDS \zi~galaxies
as a function of redshift. Fluxes are extracted from a 2 arcsec
aperture. Open and filled dots are galaxies with secure spectroscopic
redshift, respectively. Filled dots are the selected galaxies with
$M_{2000}<-19.65$. Stars are objects with tentative redshift. Crosses
have no spectroscopic redshift; in these cases we used the photometric
redshift. The selected sample contains between 25 and 38\% of the
total number of galaxies identified in GDDS with \zi~(depending
whether galaxies with tentative or photometric--redshift only are
included in the count or not).  }
\end{figure}

\section{Sample selection}

To measure metal enrichment in galaxies in the $1 <z< 2.5$ range, we
use absorption lines in the rest--frame interval $2200-2900$ \AA. This
is a particularly interesting region because it contains 7 \ion{Fe}{2}
transitions, with very different oscillator strengths, together with
the strong \ion{Mg}{2} doublet and the \ion{Mg}{1} line. The
\ion{Mg}{2} doublet is normally very strong and highly saturated but
it is very useful in determining galaxy redshifts.  The useful
wavelength range for most GDDS spectra is limited to
$\lambda=5000-9800$ \AA, or the rest--frame interval
$\lambda=2200-2900$ \AA~for galaxies at \zi.  In the total sample of
124 objects with spectroscopic redshift in the two fields, 34 galaxies
are in this $z$ range.  The number of objects with tentative or
photometric--only redshifts in \zi~is 10 and 7, respectively.

For all \zi~ objects, we estimated the AB absolute magnitude
$M_{2000}$ using a synthetic 1900 \AA~$< \lambda<$ 2100 \AA~box filter
derived from the observed $V$ and $I$ magnitudes (Fig.~1).  To
estimate $M_{2000}$ for the galaxies in our sample we use the
following empirical formula:

\begin{equation}
\begin{array}{r}
M_{2000} = V-2f(1.72)+f(z)+5.058-2.91z \\
\\
-0.114(V-I)(z-1.72)
\end{array}
\end{equation}

\noindent
where:

\begin{equation}
f(z) = 25+5\log D_L(z)-2.5\log(1+z)
\end{equation}

\noindent
and $D_L(z)$ is the luminosity distance in Mpc of galaxies at redshift
$z$.  Eqs. 1 and 2 were derived by us considering that to first order
approximation the observed $V$ band matches the rest $M_{2000}$ band
at $z=1.72$. The extra terms represent the mean $K-$correction away
from $z=1.72$ and a small correction for its dependence on spectral
energy distribution (SED) shape (parameterized by $V-I$ {\em
observed}).  The coefficients were derived by empirically fitting the
observed colors to $1.3<z<2$ template spectra (Chen et al.~2002)
ranging from E/S0 to Starburst SEDs; the formula was found to predict
$M_{2000}$ to $<0.2$ mag accuracy over this entire redshift range for
all SEDs. We emphasize that it is robust, model independent
(i.e. insensitive to SED shape) in this redshift range simply because
the optical photometry samples the rest--frame UV.

Among the 34 galaxies with secure redshifts, we selected the brightest
subsample with $M_{2000} < -19.65$ (Fig.~1). This subsample contains
17 galaxies, but three of them were not considered because the spectra
are contaminated by second order lines, and a fourth because it has
partial spectral coverage in the $2200-2900$ \AA~range. The remaining
13 galaxies (Table 1) cover the range \zii~(mean and median redshift
$z=$ \zav~and $z=$ \zme). In Table 1, we report redshifts
independently derived from the ISM absorption lines and the
[\ion{O}{2}] emission line (when observed). The mean velocity
difference between the [\ion{O}{2}] emission and ISM absorption is
$\Delta v = 93$ with 197 \kms~dispersion. The observed $I-K$ colors
(in the range $2.2<I-K<3.5$) are consistent with predictions derived
from a suite of spectral synthesis models for Irr-- or Scd--type
galaxies.  This sample represents 25\% or 38\% of the total number of
GDDS galaxies in \zi, depending whether the objects with tentative
spectroscopic or photometric--only redshifts are included in the
counting or not.

\begin{figure*}\label{f1b}
\centerline{\epsfxsize=18cm \epsffile{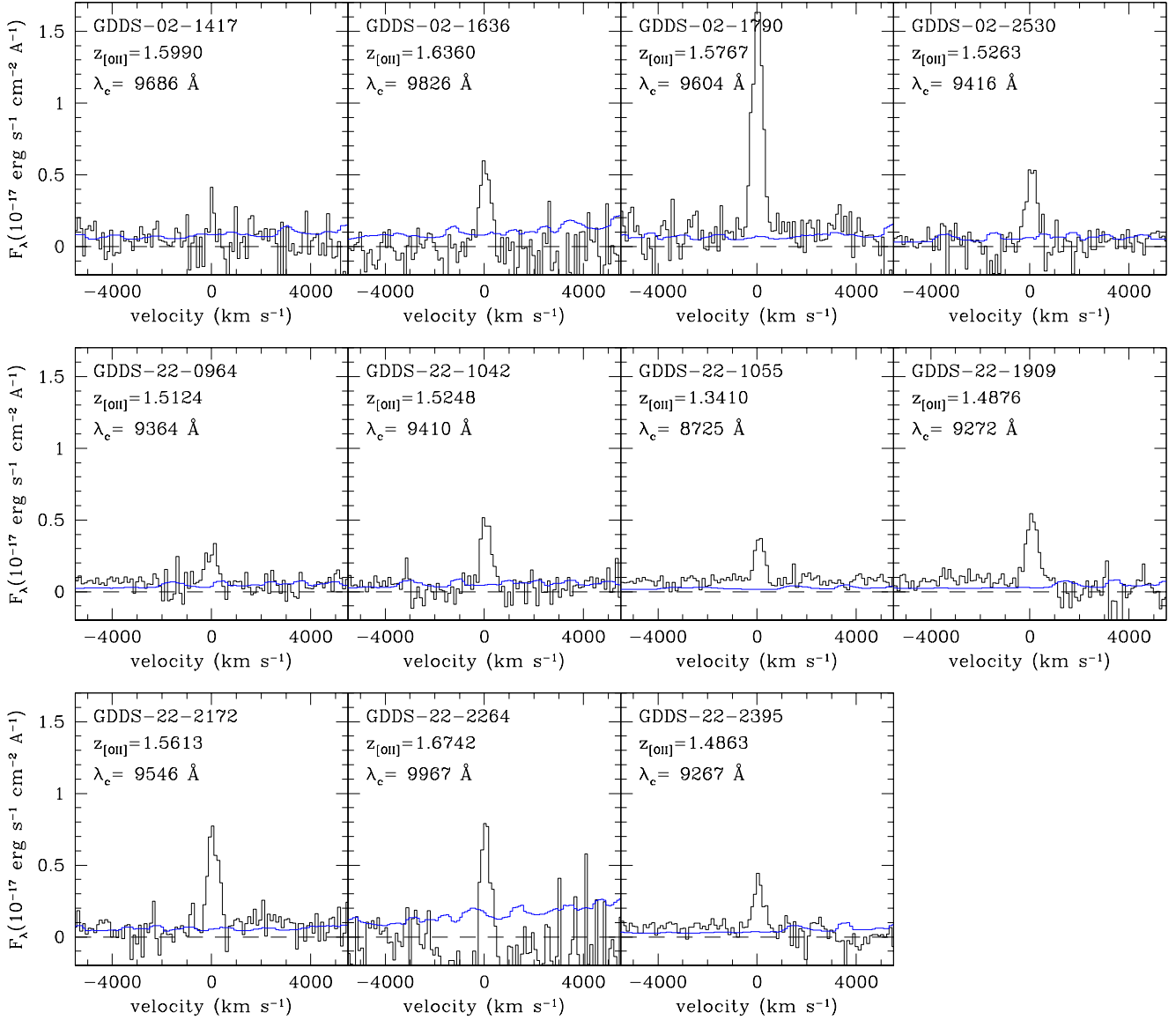}} \figcaption
{[\ion{O}{2}] emission for 11 GDDS galaxies. Smooth lines are
the noise spectra per pixel ($\sim 3.5$ \AA~in the observed
frame). Also reported is the central observed [\ion{O}{2}] emission
line wavelength $\lambda_c$.  Note that the continuum is not
subtracted from the spectra, but subtracted when measuring line
fluxes.  The flux at $\lambda \sim 9960$ \AA~in object GDDS-02-2264
goes below zero due to red--end contamination.}
\end{figure*}

\section{The [\ion{O}{2}] emission and star formation rates}

The [\ion{O}{2}] $\lambda3727$ emission line is detected in 11 of the
13 galaxies selected (Fig.~2). For the two highest redshift galaxies
the [\ion{O}{2}] wavelength is out of the observed spectral range. The
presence of the [\ion{O}{2}] nebular emission is another indication of
star formation activity in these galaxies, as it is associated with
\ion{H}{2} regions. No other emission line is detected. The
[\ion{O}{2}] flux has been measured in the spectrum of each galaxy and
reported in Table 1. As fluxes in the red end of the spectra are
roughly consistent with the aperture corrected photometry, no aperture
correction has been applied.  Errors are derived using photon counting
statistics only. Our spectral sensitivity is sufficiently good to
detect very low fluxes, down to $\sim2\times10^{-17}$ erg s$^{-1}$ \cm
~($\sim 3\sigma$) at $\lambda>8600$ \AA.  Observed fluxes, in the
range $(3-26)\times10^{-17}$ erg s$^{-1} $ cm$^{-2}$ and absolute
luminosities $(5-42)\times 10^{41}$ erg s$^{-1}$ are reported in Table
1.

The star formation rate (SFR) in local \ion{H}{2} regions is generally
directly derived from optical Balmer lines (Gallagher, Hunter, \&
Bushouse 1989; Kennicutt 1992; Gallego et al.~1995). However, at high
redshift, Balmer lines are redshifted to the IR, making the SFR
measurement harder.  If Balmer lines are not observed, the
[\ion{O}{2}] emission line can be used instead, since it is detectable
in the optical up to $z\sim 1.6$. The problem is that [\ion{O}{2}]
gives a much less direct estimate of the SFRs, because it is more
affected by dust than Balmer lines (for which dust correction can be
easily determined), and because it depends on the electron temperature and
metallicity of the \ion{H}{2} region. In local galaxies, Jansen,
Franx, \& Fabricant (2001) found a correlation between
[\ion{O}{2}]/H$\alpha$ ratio and the galaxy absolute magnitude that
allows to estimate the SFR if only the [\ion{O}{2}] emission line and
the $M_B$ are known. Their Fig.~4a shows that for bright objects
($M_B<-20.5$, as our galaxies are) H$\alpha$ is about $5-10$ times
brighter than [\ion{O}{2}]. The SFR obtained with this method is
uncertain by a factor of 5.

However, this relationship might not be valid at high redshift and the
SFR may be overestimated by up to a factor of 10 (Jansen et al.~2001;
Hicks et al.~2002). Glazebrook et al. (1999) have compared H$\alpha$
and [\ion{O}{2}] luminosities in a sample of 13 $z\sim1$ field
galaxies. The median value in the sample is $L_{\rm H\alpha} \sim
2\times L_{\rm [OII]}$; in this relation the dust correction is not
applied.

If we use the  prescription (Kennicutt 1998):

\begin{equation} 
{\rm SFR_{H\alpha} (M_\odot~ yr^{-1})} \simeq \frac{L_{\rm H\alpha}}{1.26\times10^{41} {\rm ~erg ~s^{-1}}}
\end{equation}

\noindent
that gives the SFR for a given H$\alpha$ luminosity, and assume a dust
correction (see \S 8.2) for the H$\alpha$ flux of $A_{\rm H\alpha}=
0.5$ magnitudes (also adopted by Glazebrook et al.~1999), we obtain:

\begin{equation}
{\rm SFR_ {[OII]}(M_\odot~ yr^{-1})}  \simeq \frac{L_{\rm [OII]}}{0.39\times
10^{41} {\rm ~erg ~s^{-1}}} ~,
\end{equation}

\noindent 
where the [\ion{O}{2}] luminosity is the one observed, i.e. not
corrected for dust.  From this, we derive SFRs in out sample in the
range $13-106$ M$_\odot$ yr$^{-1}$ (Table 1), with mean and median
value of $\sim 40$ M$_\odot$ yr$^{-1}$. These values are compared in
\S 8.3, with the SFRs derived using UV luminosities.

High SFR values (derived from Balmer lines) are not new to high
redshift star--forming galaxies. For instance, Glazebrook et
al. (1999) derived $20-60$ M$_\sun$ yr$^{-1}$ in their $z\sim1$
sample.  From UV fluxes, Daddi et al. (2003) derived much higher SFRs
in the $K20$ $1.7<z<2.3$ galaxies, with SFRs ranging $100-500$ M$_\sun$
yr$^{-1}$. High SFRs are also found in $2.3\leq z\leq 3.3$ LBGs, for
which infrared spectra indicate SFR $\sim20-200$ M$_\sun$ yr$^{-1}$
(Pettini et al.~1998; Kobulnicky \& Koo 2000).

\begin{figure*}\label{f1}
\centerline{\epsfxsize=18cm \epsffile{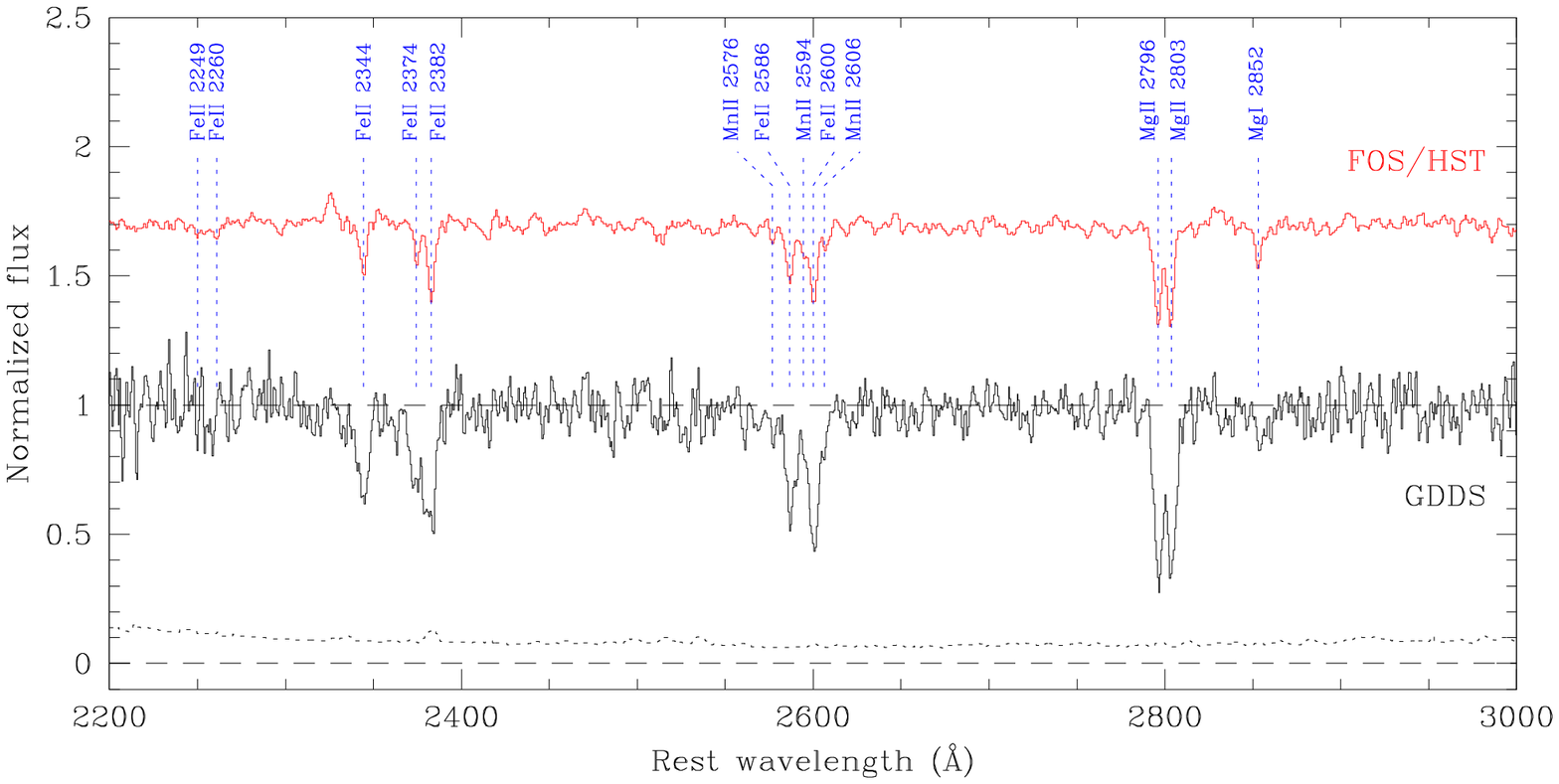}} \figcaption {Composite
spectrum of 13 GDDS galaxies with ISM absorption lines (lower
spectrum).  The noise per 0.7 \AA~pixel (dotted spectrum) is also
shown.  The sample covers the redshift range \zii~ (mean and median
redshift $z = $\zav~ and \zme, respectively).  Absorption features are
also marked. As a reference, we show the \coadd~spectrum of 14 local
starburst dwarf galaxies observed with HST/FOS (Tremonti et al.~2003),
shifted along the y--axis for comparison.}
\end{figure*}

\section{Co--addition}

In order to improve the signal--to--noise ratio in the $2200-2900$
\AA~spectral range, and to enable studies of the global properties of
the galaxy sample, the 13 selected GDDS spectra have been shifted to
the rest frame and combined. Both linearly and optimally extracted
spectra were initially used and two separate composites have been
obtained.  To combine the data in a suitable way and maximize the
signal--to--noise ratio, we have used the following procedure.  The
spectra were first normalized to one; this allowed us to analyze in
the \coadd~spectrum the EWs of lines of the same ion in a consistent
way. The continuum shape was determined by interpolating regions free
of spectral features. Then we created for each spectrum a mask
spectrum containing ``0'' or ``1'', according to whether the region or
pixel can be used or not. Generally, the spectra are relatively free
of masked pixels, however small regions in some spectra are
contaminated by strong second order refraction. Occasionally other
pixels were masked due to strong sky emission line residuals or
CCD defects.  The total number of masked pixels in the
$\lambda=2200-3000$ \AA~spectral range is 3.8\% of the total.

The observed spectra were shifted to the rest--frame (using ISM
redshifts) and linearly averaged to form a mean spectrum (i.e. each
galaxy was weighted the same in the \coadd). The \coadd~noise spectrum
is derived consistently.  We finally re--normalized the
\coadd~spectrum to adjust for misplacements of the continuum in the
individual spectra. The uncertainty in the continuum of the \coadd~is
estimated to be $\sim 3-5$\%, from the higher to the lower signal
regions, and is consistent with the level of the noise per resolution
element.

The \coadd~optimal spectrum is shown in Fig.~3 (lower spectrum). For
comparison, we also show (shifted upwards along the $y-$axis) the
normalized \coadd~of 14 spectra of local starburst galaxies (Tremonti
et al. 2003) obtained with the low resolution Faint Object
Spectrograph (FOS) on board of the Hubble Space Telescope (HST). The
GDDS mean FWHM in the rest frame is 6.1 \AA, and is estimated by
considering the observed FWHM divided by $(1+\langle z \rangle)$,
where $\langle z \rangle =$ \zav~ is the mean redshift of the sample.
For the \coadd~we have chosen a small pixel size of 0.7 \AA~($\Delta
v\simeq 80$ \kms~at 2600 \AA), i.e.~$\sim2$ times smaller than the
real mean pixel size (1.39 \AA). This allows us to better identify
small features in the spectrum and more accurately measure line
EWs. The resolution element is extended over $\sim9$ pixels.  The
\coadd~optimal noise spectrum per 0.7 \AA~pixel is also shown as a
dotted line in Fig.~3.

As a consistency check, the EWs measured from the optimal and linear
\coadd~spectra were compared. The agreement is very good and no
correlation with, e.g., wavelength is present.  Since the
signal--to--noise of the optimally extracted \coadd~is 5--15\% higher
than the linearly extracted one, the optimal \coadd~is used for all
subsequent analysis.

The mean S/N per 0.7 \AA~pixel in regions free of strong features has
been estimated using the noise spectrum. This is S/N = 9.1 at $\lambda
= 2210-2330$ \AA, S/N = 12.3 at $\lambda=2395-2570$ \AA, S/N = 14.2 at
$\lambda=2615-2785$ \AA, S/N = 11.8 at $\lambda=2815-2990$ \AA. The
S/N measured with the noise spectrum rescales with the square-root of
the pixel size chosen.  As a sanity check, we also measured the
standard deviation directly in the signal spectrum.  We obtain S/N =
11.2, 13.6, 18.3, and 13.9, in the same regions as before, i.e. about
1.2 times higher than that obtained from the noise spectrum. This is
close to the expected value, $\sqrt{2}\simeq 1.4$, given that the
chosen pixel size (0.7 \AA) is $\sim2$ smaller than the real pixel
size (1.39 \AA).

\begin{table}
\caption[t1]{Rest--frame equivalent widths}\label{t1}
\begin{center} 
\begin{tabular}{lcccc} 
\tableline\tableline&&&&\\[-5pt] 
&& \multicolumn{3}{c}{$W_r$ (\AA) } \\
[4pt]\cline{3-5}\\[-4pt] 
Line     & $f_\lambda$ &  GDDS   & cB58 \tablenotemark{a} & LBGs \tablenotemark{b} \\ 
[5pt]\tableline&&&&\\[-5pt] 
\ion{Si}{2} $\lambda1260$ & 1.0070  & $\cdot\cdot\cdot$ & $(2.54\pm0.07)$ & $<1.7$ \\
\ion{Si}{2} $\lambda1526$ & 0.1270  & $\cdot\cdot\cdot$ & $2.59\pm0.03$ & $1.72\pm0.18$ \\
\ion{Si}{2} $\lambda1808$ & 0.00218 & $\cdot\cdot\cdot$ & $0.53\pm0.02$ & $(0.04-0.28)$ \\  
\ion{Fe}{2} $\lambda1608$ & 0.058   & $(1.62\pm0.12)$ & $1.30\pm0.04$ & $0.91\pm0.15$ \\
\ion{Fe}{2} $\lambda2249$ & 0.00182 & $<1.2$        & $(0.15\pm0.01)$& ($0.048-0.13$) \\
\ion{Fe}{2} $\lambda2260$ & 0.00244 & $<0.9$        & $(0.19\pm0.01)$& $(0.065-0.18)$ \\
\ion{Fe}{2} $\lambda2344$ & 0.114   & $2.90\pm0.27$ & $2.99\pm0.04$ &  $\cdot\cdot\cdot$ \\
\ion{Fe}{2} $\lambda2374$ & 0.0313  & $2.28\pm0.23$ & 1.94\tablenotemark{c}& $(0.9-1.4)$ \\
\ion{Fe}{2} $\lambda2382$ & 0.320   & $3.57\pm0.31$ & 3.41\tablenotemark{d}&$\cdot\cdot\cdot$  \\
\ion{Fe}{2} $\lambda2586$ & 0.0691  & $2.96\pm0.17$ & $(2.88\pm0.06)$& $\cdot\cdot\cdot$ \\
\ion{Fe}{2} $\lambda2600$ & 0.239   & $3.94\pm0.27$ & $(3.95\pm0.10)$& $\cdot\cdot\cdot$ \\
\ion{Mn}{2} $\lambda2576$ & 0.3508  & $0.56\pm0.13$ & $0.38\pm0.10$ & $\cdot\cdot\cdot$ \\
\ion{Mn}{2} $\lambda2594$ & 0.2710  & $(0.45\pm0.16)$ & ($0.30\pm0.09$) & $\cdot\cdot\cdot$ \\
\ion{Mn}{2} $\lambda2606$ & 0.1927  & $(0.33\pm0.12)$ & ($0.22\pm0.06$) & $\cdot\cdot\cdot$ \\   
\ion{Mg}{2} $\lambda2796$ & 0.6123  & $4.10\pm0.19$ & $\cdot\cdot\cdot$ & $\cdot\cdot\cdot$ \\
\ion{Mg}{2} $\lambda2803$ & 0.3054  & $3.98\pm0.16$ & $\cdot\cdot\cdot$ & $\cdot\cdot\cdot$ \\
\ion{Mg}{1} $\lambda2852$ & 1.8100  & $0.91\pm0.19$ & $\cdot\cdot\cdot$ & $\cdot\cdot\cdot$ \\
[2pt]\tableline
\end{tabular}
\tablenotetext{}{{\sc Note.} -- Oscillator strengths $f_{\lambda}$ are taken from a
recent compilation by Prochaska et al.~(2001). EWs between brackets
are extrapolated from measured column densities (see text). EW upper
limits for \ion{Fe}{2} in GDDS \coadd~are $4\sigma$.}
\tablenotetext{a}{EWs taken from Pettini et al.~(2002).}  
\tablenotetext{b}{EWs taken from Shapley et al.~(2003).}
\tablenotetext{c}{The $1\sigma$ error reported by Pettini et al.~(2002) is $>0.07$ \AA~(line is affected by bad pixels).}
\tablenotetext{d}{The $1\sigma$ error reported by Pettini et al.~(2002) is $>0.05$ \AA~(line is affected by bad pixels).}
\end{center}
\end{table}

\section{Heavy element enrichment}

The ultimate goal of the present analysis is to estimate the heavy
element enrichment of galaxies in the GDDS composite spectrum. In the
observed range, we have clearly detected \ion{Fe}{2}, \ion{Mg}{2} and
\ion{Mg}{1} absorption lines (Fig.~3). Besides these, we have also
included in the analysis the very weak \ion{Mn}{2} triplet around 2600
\AA.  All EWs with errors in the GDDS \coadd~are reported in the third
column of Table 2. The $4\sigma$ EW upper limits for two \ion{Fe}{2}
non--detections are also estimated using the noise spectrum, by
assuming that lines are extended over 6 \AA~(or $\Delta v \sim 800$
\kms~at $\lambda\sim2250$ \AA).  Values in brackets are EWs obtained
by extrapolating from estimated column densities, see \S~\ref{s3.1}.
To check the line de--blending in the \ion{Fe}{2}
$\lambda\lambda2374,2382$ and \ion{Mg}{2} $\lambda\lambda2796,2803$
doublets, we compared the measured EWs in the spectrum with the EWs
given by the fit of two blended Gaussians. For the \ion{Fe}{2} doublet
we measure a ratio between the two lines of 0.59, to be compared with
0.64 given by the ratio of the two Gaussians.  For the \ion{Mg}{2}
doublet we obtain 1.05, to be compared with 1.03 given by the two
Gaussians. This shows that the contamination is properly accounted
for. The contamination of the \ion{Fe}{2} $\lambda2600$ line by
\ion{Mn}{2} is treated separately (see next section).

The UV continuum of star--forming galaxies (like our selected sample)
is dominated by O and B stars. In this case, the stellar contamination
to the ISM lines in the $2200-3000$ \AA~interval is negligible
(Tremonti et al.~2003). Moreover, when stellar wind is not important
(this is the case for lines in the $2000-3000$ \AA~region) the
strength of the photospheric lines is significantly weaker when
compared to ISM lines (de Mello, Leitherer, \& Heckman 2000).

The lines detected in the GDDS \coadd~are strong, but narrow in
velocity space ($\Delta v < 800$ \kms). Assuming that these
interstellar lines originate in gas viewed against the continuum
provided by the integrated light of O and B stars in the galaxies, we
can measure a sort of ``average'' column densities of the ions. Two
approaches have been followed. In the first case we used EWs as given
in Table 2 with the curve of growth analysis (COG, Spitzer 1978).  In
the second, we simultaneously fit absorption profiles to all lines in
the spectrum arising from transitions of the same ion. The advantage
of fitting line profiles is that line blending is automatically taken
into account.

\begin{figure}\label{f2}
\centerline{{\epsfxsize=9cm \epsfbox{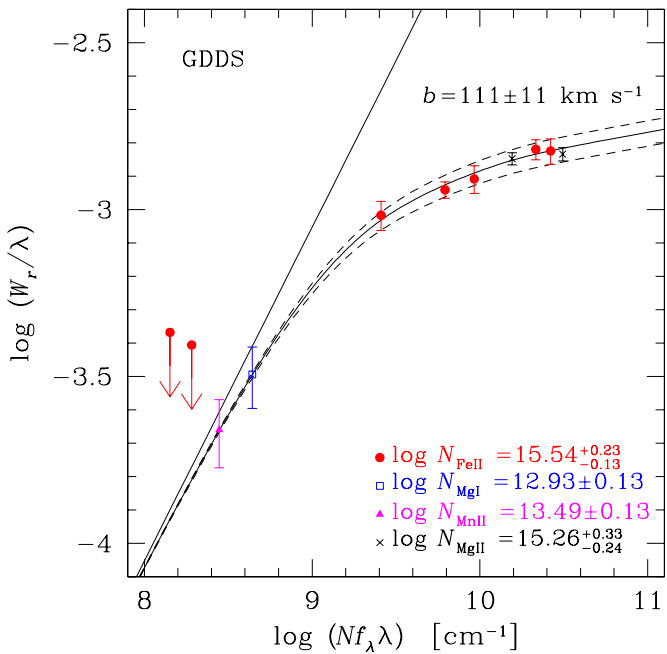}}}
\figcaption[f2]{Best fit curve of growth for \ion{Fe}{2} in the GDDS
composite spectrum. The solid and dashed lines are for the best
fit Doppler parameter and $\pm1\sigma$, respectively. The other ion
column densities are calculated assuming the best fit $b$ value found
for \ion{Fe}{2}.  The straight line shows the linear approximation to the
curve of growth ($b=+\infty$).}
\end{figure}

\begin{figure}\label{f2d}
\centerline{{\epsfxsize=9cm \epsfbox{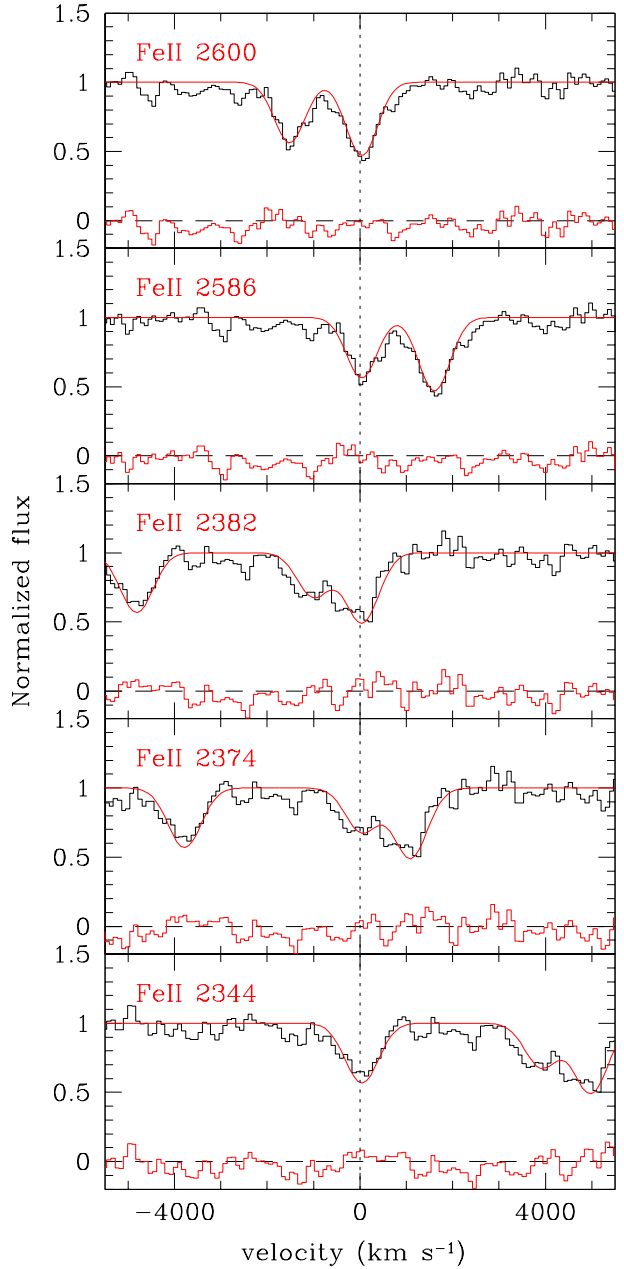}}}
\figcaption[f1]{Line absorption fitting profile of \ion{Fe}{2} for
GDDS \coadd.  The smooth line is the best fit with
\feii~$=15.52^{+0.15}_{-0.12}$ and $b=115\pm10$ \kms. Shown at the
bottom are the best fit residuals.}
\end{figure}

\subsection{Column densities}\label{s3.1}

The rest--frame EWs of the ISM lines detected in the GDDS
\coadd~galaxy are very high, indicating that the gas is heavily
enriched by metals. As the majority of absorption lines are very
strong ($W_r>1$ \AA), they deviate from the linear part of the COG and
it is not possible to use the linear approximation to determine column
densities. Moreover, the low resolution (FWHM $\sim 700$ \kms~ at
$\lambda \sim 2600$ \AA) does not allow application of the apparent
optical depth method (Savage \& Sembach 1991).  However, even with the
relatively large uncertainties, the general COG analysis can still
give significant results since so many lines are present.  Given that
several galaxy spectra are combined together, it is not possible to
separate the different components of very complex absorption features
and therefore they appear as single lines (Fig.~3).  Hence, the widths
of these features do not provide useful information on the kinematics
or thermal conditions of the gas in the galaxies and the broadening is
a sort of ``effective'' Doppler parameter, being the result of the
superposition of many absorbing clouds.  The COG analysis with a
single component approximation was used for QSO--DLA studies
before high resolution spectroscopy of QSOs became possible (see for
instance Blades et al.~1982; Turnshek et al.~1989; Steidel 1990). In
fact, Jenkins (1986) showed that the COG technique applied to complex
features nearly always gives a reasonable answer (the
simulated--to--true column density ratio rarely goes below 0.8) even
if different lines have very different saturation levels or Doppler
parameters.  The general results have been confirmed by Savage \&
Sembach (1991), who in addition, show that the COG works best if the
gas absorption is not ``bimodal'' -- e.g., that very large absorptions
are not combined with narrow ones.  Since our composite spectrum is
derived from many galaxies, each with many absorption clouds, a
bimodal distribution of absorption lines is very unlikely.

In Fig.~4 we show the COG for the absorption lines of the GDDS
composite spectrum. The fit to the data was determined by $\chi^2$
minimization of the COG. The resulting column densities are reported
in Table 3. To determine the \ion{Fe}{2} column density, we have used
5 strong detections, and iterated the results with those for
\ion{Mn}{2}, because two of the three weak \ion{Mn}{2} lines are
blended with the \ion{Fe}{2} $\lambda2600$ line. The weakest
\ion{Fe}{2} line detected, \ion{Fe}{2} $\lambda 2374$, gives $\log
N_{\rm FeII}>15.2$, using the linear approximation for the COG.  The
non--detection of \ion{Fe}{2} $\lambda2260$ line sets an upper limit
of $\log N_{\rm FeII} < 16$. These two limits already give a good,
robust estimate of the \ion{Fe}{2} column density. The 5 \ion{Fe}{2}
EWs give a best fit column density of \feiib, and an effective Doppler
parameter $b=111\pm11$ \kms. From this, we extrapolate the EW of the
\ion{Fe}{2} $\lambda1608$ line to be $W_r = 1.62\pm0.12$ \AA. The
reduced $\chi^2$ value obtained from the best fit is 0.34, indicating
that the errors are probably not underestimated.  The \ion{Mn}{2}
$\lambda2576$ line is isolated and marginally detected with
$4.4\sigma$ significance (Table 2). Moreover, the presence of the
\ion{Mn}{2} absorption is supported by the large EW of the \ion{Fe}{2}
$\lambda2600$ line, indicating the presence of \ion{Mn}{2}
$\lambda\lambda2594,2606$ there too.  The column density derived from
the \ion{Mn}{2} $\lambda2576$ line is in the range $13.25 <$
\mnii~$<13.67$, assuming $b>90$ \kms, with very little
saturation. Taking the same $b$ value as found for \ion{Fe}{2}, we
find a best fit \mniib. From this, we derive $W_r = 0.45\pm0.12$
\AA~and $0.33\pm0.10$ \AA, for \ion{Mn}{2} $\lambda2594$ and
\ion{Mn}{2} $\lambda2606$, respectively. These two values have been
subtracted from the \ion{Fe}{2} $\lambda2600$ EW and the error
propagated.  The weak \ion{Mg}{1} $\lambda2852$ line gives \mgib~ for
$b=111$ \kms.  The \ion{Mg}{2} doublet is heavily saturated, namely
\mgii~$>14.3$. We estimated \mgiib~for a Doppler parameter
$b=111\pm11$ \kms~(as for the \ion{Fe}{2} best fit).

As a further test on the \ion{Fe}{2} column density
measurement, we have performed simultaneous line profile fitting of
the 5 lines. The result with one component gives
\feii~$=15.52^{+0.15}_{-0.12}$ and $b=115\pm10$ \kms~(Fig.~5), very
similar to that found with the COG method (\feiib), with a reduced
$\chi^2=0.55$.  This shows consistency between the line profiles and
the column density derived using the COG. The velocity shift between
the fit and the rest frame is consistent with zero.

We finally considered another source of uncertainty: the ISM covering
factor. So far we assumed that the covering factor is one, that is the
gas is uniformly distributed in front of the emitting sources, and
photons escaping the absorbing gas or scattered are not important.
However, in case of a non--unity covering factor, the residual flux
not absorbed by the ISM would cause a systematic underestimate of the
column densities, if not properly taken into account.  From the depth
of the MgII absorption, we have estimated that the covering factor is
at least 70\%. We recalculated the FeII column density, assuming this
very conservative lower limit, and obtained $N_{\rm FeII} \sim 60$\%
higher than in the case of unity covering factor.  In log scale the
difference is only 0.2 dex: $\log N_{\rm FeII} = 15.75$ instead of
$\log N_{\rm FeII}=15.54$; this is within the upper error estimated in
our first measurement (0.23 dex).  Shapley et al. (2003) discuss the
non--unity of the covering factor in LBGs. However, Giallongo et
al.~(2002) found that in two LBGs the escaping fraction of ionizing
photons ($\lambda<912$ \AA) is $<16$\%, whereas Malkan, Webb \&
Konopacky (2003) found that the same quantity in the UV images of
11 bright blue $1.1<z<1.4$ galaxies is less than $6$\%. Both results
evoke some caution on this issue.

\section{Metal column densities in high redshift ISM}

When the gas is mostly neutral ($N_{\rm HI} \sim N_{\rm H}$), the
column density of an element can be approximated with the column
density of the ion with ionization potential above the hydrogen
ionization potential (13.6 eV). For instance, iron, magnesium and
manganese are mainly in the Fe$^+$, Mg$^+$, and Mn$^+$ state. In the
GDDS \coadd, the very high EWs of \ion{Fe}{2} and \ion{Mg}{2}
indicates that the gas is nearly neutral and the $N_{\rm FeII} \sim
N_{\rm Fe}$, $N_{\rm MgII} \sim N_{\rm Mg}$, or $N_{\rm MnII} \sim
N_{\rm Mn}$ approximation can be applied.  The low ionization of the
gas is confirmed by the presence of a relatively strong \ion{Mg}{1}
absorption (\mgi~$\sim12.9$), but with column density much lower than
the \ion{Mg}{2} column density (\mgii~$\sim15-15.6$).

Considering column densities of these ions in the neutral ISM instead
of metallicities (element-to-hydrogen relative abundance) has the
advantage that we can directly compare results with other classes of
objects at various redshifts with similar measurements, without
knowing the \ion{H}{1} content. Indeed, \ion{H}{1} absorption in
GDDS spectra cannot be detected for lack of spectral coverage.  In
Table 3 we compare column densities of \ion{Fe}{2}, \ion{Mn}{2},
\ion{Si}{2}, observed in the GDDS \coadd, and other high redshift and
local galaxies.

Heavy element column densities in the neutral ISM of high redshift
galaxies have been measured in great detail for more than 100
QSO--DLAs in the interval $0.0 < z < 4.5$ (see for instance Lu et
al.~1996; Pettini et al.~1997; Prochaska \& Wolfe 1999; Prochaska et
al.~2003a).  For our purpose, we have used our own large compilation
of DLAs\footnote{For the largest public compilation of DLA column
densities, visit J.~Prochaska's homepage at
http://www.ucolick.org/$\sim$xavier/ESI.}. The mean column densities
for \ion{Fe}{2}, \ion{Mn}{2}, and \ion{Si}{2} with dispersions are
given in Table 3. We have divided the sample in subsamples for each
ion. The number of DLAs considered in each subsample is also given in
the first column (between brackets). The column density distributions
for \ion{Fe}{2} and \ion{Mn}{2} as a function of redshift are shown in
Figs.~7 and 8.

The heavy element column densities in the ISM of GRB--DLAs have
been studied in only 4 objects (Savaglio et al.~2003, Vreeswijk et
al.~2003).  In a fifth event, GRB~020813 ($z=1.255$), equivalent
widths of many UV lines have been reported by Barth et al.~(2003),
from which we have estimated column densities using the COG (Table 3).

We also include in our analysis, the ISM lines measured in
star--forming galaxies in the local Universe observed with the Far
Ultraviolet Spectroscopic Explorer (FUSE). The sensitivity and
resolution (FWHM $\sim 30$ \kms) of FUSE is good enough for a detailed
analysis of nearby starburst galaxies. In our study we considered
metal column densities obtained for NGC 1705 (Heckman et al.~2001) and
I~Zw~18 (Aloisi et al.~2003).  For comparison, we have also included
measurements in the diffuse and dense ISM of the Milky Way and the
Magellanic Clouds (Sembach \& Savage 1996; Spitzer \& Fitzpatrick
1995; Welty et al.~1997, 1999).

The neutral ISM metal enrichment of LBGs has been discussed by Shapley
et al.~(2003) for a composite of 811 spectra, and by Pettini et
al.~(2002) for the bright LBG galaxy cB58 at $z=2.72$. In the
following sections, we use their EWs and estimate heavy element column
densities using the COG.

\begin{figure}\label{f2b}
\centerline{{\epsfxsize=9cm \epsfbox{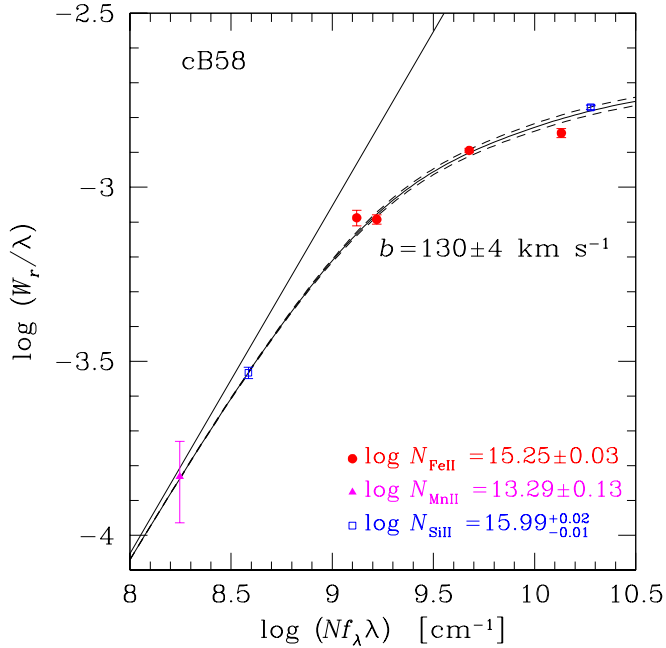}}}
\figcaption[f1]{Best fit curve of growth for \ion{Fe}{2} in the cB58
ISM. The solid and dashed lines are for the best fit Doppler parameter
and $\pm1\sigma$, respectively. The \ion{Mn}{2} column density is
calculated assuming the best fit $b$ value found for \ion{Fe}{2}. The
best fit for \ion{Si}{2} is calculated independently and is not shown
for clarity, but this is very similar to the \ion{Fe}{2} best fit. The
straight line shows the linear approximation to the curve of growth
($b=+\infty$).}
\end{figure}

\subsection{The spectrum of the LBG MS 1512--cB58}\label{s4.2}

Column densities in the LBG MS 1512--cB58 ($z=2.72$) have been
determined by Pettini et al.~(2002) using the apparent optical depth
(Savage \& Sembach 1991) measured on the less saturated lines.  For a
comparison of the COG with this method, we have redone the analysis
for cB58 using the COG.  The EWs used are given in the fourth column
of Table 2 and are the ones reported by Pettini et al.~(2002).  The
COG for cB58 is shown in Fig.~6 and the column densities are listed in
Table 3. The EWs of two of the four \ion{Fe}{2} lines, \ion{Fe}{2}
$\lambda\lambda2374,2382$, have errors larger than 0.07 and 0.05
\AA~for contamination by bad pixels (Pettini et al.~2002), for which
we arbitrarily use $\sigma=0.15$ \AA.

The advantage of using the COG is that all detected \ion{Fe}{2} lines
can be used, regardless of the saturation level.  The apparent optical
depth method, as Pettini et al.~(2002) points out, can still
underestimate column densities due to saturation.  This effect is also
noted in the ISM of local dwarf galaxy I Zw 18 by Aloisi et
al.~(2003), who made a comparison between the line fitting of Voigt
profiles and the apparent optical depth method.  The COG analysis for
cB58 is shown in Fig.~6.  The best fit to the four \ion{Fe}{2} lines
gives \feii~$15.25^{+0.04}_{-0.03}$ and $b=130\pm4$
\kms~($\chi^2=3.9$), consistent with \feii~$\simeq15.20$ found by
Pettini et al.~(2002). Errors on column densities derived by Pettini
et al.~(2002) are reported to be generally less than 0.2 dex.

The \ion{Mn}{2} $\lambda 2576$ line is very close to the linear part of the
COG, therefore the column density is relatively well constrained
in the range \mnii~$=13.12-13.44$ for $b>90$ \kms. If we adopt $b=130$
\kms, as given by \ion{Fe}{2}, we obtain \mnii~$=13.29\pm0.13$, consistent
with \mnii~$\simeq13.33$ found by Pettini et al.~(2002).  From this
column density and Doppler parameter, we derived the expected EWs of
the \ion{Mn}{2} $\lambda\lambda2594,2607$ doublet (Table 2).

For completeness, we also estimate the \ion{Si}{2} column density.
\ion{Si}{2} $\lambda1260$ is weakly contaminated by \ion{S}{2}
$\lambda1259$ and \ion{Fe}{2} $\lambda1260$ ($W_r\sim1$ \AA),
therefore we do not use it.  The other two detected lines have very
different oscillator strengths. The \ion{Si}{2} $\lambda1808$ line is
weak and little saturated, and gives a column density in the range
$15.92< \log N_{\rm SiII} < 16.05$ for $b>70$ \kms. From \ion{Si}{2}
$\lambda\lambda1526,1808$ together, we derive
\siii~$=15.99_{-0.02}^{+0.01}$ and $b=136\pm2$ \kms, consistent with
$\log N_{\rm SiII}\simeq 15.99$ given by Pettini et al.~(2002). We
also note that the effective Doppler parameter for \ion{Fe}{2} and
\ion{Si}{2} have been found independently to be very similar ($b_{\rm
FeII}=130$ \kms~and $b_{\rm SiII}=136$ \kms).

We found an excellent agreement between the results obtained by
Pettini et al. (2002) using the apparent optical depth method and ours
obtained with the COG (Table 3), supporting the robustness of our
results for GDDS galaxies.

\begin{table*}
\caption[t2]{Interstellar column densities}\label{t2} 
\begin{center} 
\begin{tabular}{lcccccc}
\tableline\tableline&&&&&&\\[-5pt]
&& \multicolumn{5}{c}{$\log N$ (\cm) } \\
[4pt]\cline{3-6}\\[-4pt] 
Galaxy & redshift &\ion{Fe}{2} & \ion{Mn}{2} & \ion{Si}{2} & \ion{H}{1} & Ref \\
[2pt]\tableline&&&&&&\\[-5pt]
GDDS (13) & \zav   & $15.54^{+0.23}_{-0.13}$ & $13.49\pm0.13$ & $\cdot\cdot\cdot$ & $\cdot\cdot\cdot$ & 1 \\
MW--MC    & 0.00   & $\langle 14.89 \rangle$ &  $\langle 13.00 \rangle$   & $\langle 15.53 \rangle$ & $\cdot\cdot\cdot$ & 2,3,4 \\
I Zw 18   & 0.0026 & $15.09\pm0.06$          & $\cdot\cdot\cdot$    & $14.81\pm0.07$ & $21.34\pm0.11$ &  5 \\
NGC 1705  & 0.0021& $14.54\pm0.10$          & $\cdot\cdot\cdot$    & $14.61\pm0.14$  & $20.2\pm0.2$   &  6 \\
GRB~990123 & 1.6004 & $14.78^{+0.17}_{-0.10}$ & $\cdot\cdot\cdot$    & $\cdot\cdot\cdot$  & $\cdot\cdot\cdot$ &  7 \\ 
GRB~000926\tablenotemark{a} & 2.038  & $15.60^{+0.20}_{-0.15}$ & $\cdot\cdot\cdot$    & $16.47^{+0.10}_{-0.15}$ & $\sim 21.3$ & 7 \\
GRB~010222 & 1.475  & $15.32^{+0.15}_{-0.10}$ & $13.61^{+0.08}_{-0.06}$ & $16.09\pm0.05$       & $\cdot\cdot\cdot$ &  7 \\
GRB~020813\tablenotemark{b} & 1.255  & $15.52^{+0.02}_{-0.03}$ & $13.63\pm0.02$ & $16.28\pm0.03$ & $\cdot\cdot\cdot$ & 1 \\
GRB~030323 & 3.3714 & $15.93\pm0.08$ & $\cdot\cdot\cdot$  & $\cdot\cdot\cdot$ & $21.90\pm0.07$ & 8 \\
cB58\tablenotemark{c}& 2.72   & $15.25^{+0.04}_{-0.03}$ & $13.29\pm0.13$       & $15.99^{+0.01}_{-0.02}$ & $20.85\pm0.09$ & 9 \\
LBGs (811)\tablenotemark{c}& $\sim3$& $14.8-15.2$           & $\cdot\cdot\cdot$    & $14.8-15.7$  & $\cdot\cdot\cdot$ & 10 \\
QSO--DLAs (120)\tablenotemark{d}~~ & $0.56-4.47$ & 14.39 (0.62)\tablenotemark{e} & $\cdot\cdot\cdot$    & $\cdot\cdot\cdot$  & 20.60 (0.51)\tablenotemark{e} & 11 \\
QSO--DLAs (21)\tablenotemark{d}~~  & $0.56-2.78$ & $\cdot\cdot\cdot$  &  12.68 (0.44)\tablenotemark{e} & $\cdot\cdot\cdot$   & 20.68 (0.56)\tablenotemark{e} & 11 \\
QSO--DLAs (105)\tablenotemark{d}~~ & $0.68-4.47$ & $\cdot\cdot\cdot$  & $\cdot\cdot\cdot$    &  14.71 (0.70)\tablenotemark{e} & 20.52 (0.50)\tablenotemark{e} & 11 \\
[2pt]\tableline\\
\end{tabular}
\vskip 7pt
\begin{minipage}{11cm}
\tablenotetext{}{References. -- (1) This  work; (2) Sembach \& Savage (1996); (3) Spitzer 
\& Fitzpatrick (1995); (4) Welty et al.~(1997, 1999);
(5) Aloisi  et al.~(2003);  (6) Heckman  et al.~(2001); (7)  Savaglio et
al. (2003); (8) Vreeswijk et al.~2003; (9) Pettini et al.~(2002); (10) Shapley et al.~(2003); (11) Obtained from a compilation of QSO--DLAs}
\tablenotetext{a}{\ion{H}{1} column density estimated by Fynbo et al.~2002.}
\tablenotetext{b}{Line EWs are taken from Barth et al. 2003.}
\tablenotetext{c}{We have estimated column densities using the COG.}
\tablenotetext{d}{In brackets is the number of QSO--DLAs used.}
\tablenotetext{e}{In brackets is the dispersion in the sample.}
\end{minipage}
\end{center}
\end{table*}

\subsection{The LBG composite spectrum}

Shapley et al.~(2003) have used 811 LBG spectra in the redshift range
$2<z<3.8$ to construct a UV ($\lambda=950-1850$ \AA) composite
spectrum. Their composite is rather complex, showing stellar features,
low and high ionization absorption, and nebular emission lines. EWs of
ISM lines are reported, but column densities are not derived because
lines are heavily saturated.  They also subdivided the sample into 4
subsamples according to the Ly$\alpha$ equivalent width, in order to
find possible correlation with observed parameters. Here we only
consider the EWs of the total composite. Although the few ISM lines
are saturated ($W_r>0.9$ \AA), inspecting them with the COG can
provide interesting, even if qualitative, results on the heavy
element enrichment in the neutral ISM. We assume that the covering
factor of the ISM in front of the UV sources is one. If the covering
factor is less than one, column densities derived using the
COG would be systematically lower than real values.

Only \ion{Fe}{2} $\lambda1608$ is measured for \ion{Fe}{2}, but
because this line is moderately saturated, it can be used to
constraint the column density. We derive $14.76< \log N_{\rm
FeII}<15.21$, assuming an effective Doppler parameter $b>90$ \kms~(a
reasonable assumption given that 811 galaxy spectra are coadded). From
this, we derive the expected EWs for other \ion{Fe}{2} transitions
(between brackets in Table 2).  The \ion{Si}{2} $\lambda1304$ line is
contaminated by \ion{O}{1} $\lambda1302$, while \ion{Si}{2}
$\lambda1260$ is contaminated by \ion{S}{2} $\lambda1259$ and
\ion{Fe}{2} $\lambda1260$. To determine the \ion{Si}{2} column
density, we used these two EWs as upper limits and in addition the
\ion{Si}{2} $\lambda1526$ measurement. Assuming $b>90$ \kms, we get
\siii~$=14.8-15.7$, from which we expect $W_r$(\ion{Si}{2}
$\lambda1808)=0.04-0.28$ \AA~(Table 2).

A comparison between the GDDS composite and the LBG composite can be
made by considering the \ion{Fe}{2} $\lambda1608$ line.  Given the
good accuracy of our \ion{Fe}{2} column density measurement, we can
estimate the expected EW for the \ion{Fe}{2} $\lambda1608$ line (that
is not covered by our data), and get $W_r = 1.6\pm0.2$ \AA~(Table
2). This is 1.8 times higher than the EW of the same line measured by
Shapley et al.~(2003) in the LBG composite ($W_r=0.91\pm0.15$ \AA),
but it is similar to that of the LBG subset (1/3 of the galaxies) that
does not display Ly$\alpha$ emission ($W_r = 1.57\pm0.21$ \AA).

\section{Discussion}

Various methods have been traditionally used to derive metallicities
in high redshift galaxies using both absorption and emission
lines. The Lick indices method (Faber et al. 1985), recently applied
on SDSS $0.15 < z<0.50$ galaxies (Eisenstein et al.~2003), compares
absorption line strengths in stars with models (Vazdekis 1999) and is
mostly sensitive to old stellar population systems. As Eisenstein et
al.~(2003) point out, the models are limited in the range of
metallicities and can suffer from the age--metallicity degeneracy.

Nebular emission in the ionized ISM is, on the other hand, more
appropriate when dealing with star--forming galaxies.  The
metallicity, measured using [\ion{O}{2}] $\lambda3727$, [\ion{O}{3}]
$\lambda\lambda4959,5007$, and Balmer emission lines of 56 galaxies at
$0.26 <z<0.86$, ranges from half to two times solar (Kobulnicky et
al. 2003). In a similar study (66 galaxies at $0.5 < z < 0.9$) Lilly,
Carollo \& Stockman (2003) find that most galaxies in their sample
have $Z\sim Z_\odot$.  This method relies on a good calibration of
metallicity vs. line strengths derived from samples in the local
Universe, and requires also a reliable estimate of the stellar Balmer
absorption line contamination. As Kobulnicky et al.~(2003) mention, it
is useful in a statistical sense when using large data sets. Moreover,
this method requires IR spectroscopy for galaxies at $z>0.8$.

In our first analysis of GDDS data, we used UV absorption lines
associated with the neutral ISM to evaluate heavy element enrichment
in a subset of galaxies in \zii.  This sample represents a joint set
of the most massive galaxies at $z>1.3$ (from the deep $K$ selection)
together with those with the strongest rest--frame UV emission (from
the $M_{2000}$ selection). Thus we are selecting massive starburst
galaxies; preliminary stellar mass estimates (Glazebrook et al.~in
preparation) indicate that they have masses $M\gsim 10^{10}$ M$_\odot$
and their descendants must correspond to galaxies of at least this
mass today. This selection is quite similar to that of Lilly et
al.~(2003) at $z\sim 0.7$ whose primary $I$-band selection pulls out
massive galaxies and whose emission line cut pulls out a high--SFR
subset.

\begin{figure}\label{f3}
\centerline{{\epsfxsize=9cm \epsfbox{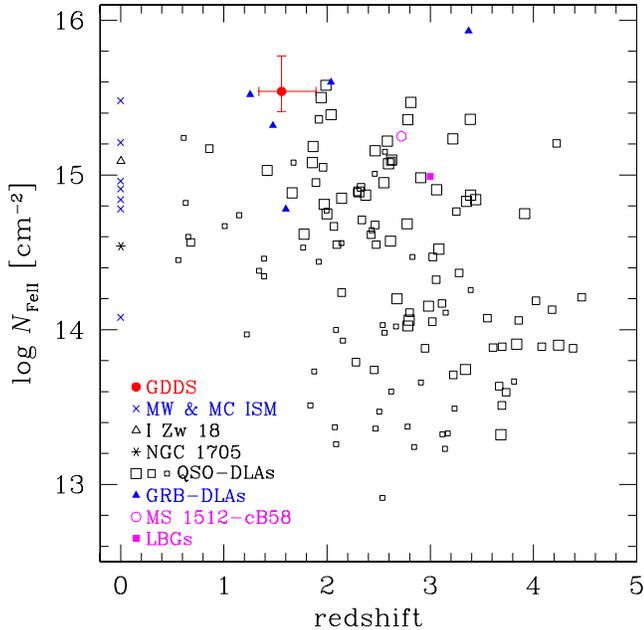}}}
\figcaption[f1]{\ion{Fe}{2} column densities as a function of
redshifts in different galaxies. QSO--DLAs are divided in three
subsamples where small, medium and big squares indicate $\log N_{\rm
HI}<20.45$, $20.45< \log N_{\rm HI} <20.80$, and $\log N_{\rm HI}
>20.80$ systems, respectively.  Errors for \feii~in QSO--DLAs is
always $<0.2$ dex, and on average is $0.05$ dex. Errors for local ISM
column densities, when available, are typically less than 0.1 dex.
For \feii~in the LBGs, we have taken the middle point in the interval
\feii~$=14.76-15.21$. Errors on the other column densities are given
in Table \ref{t2}.}
\end{figure}

\begin{figure}\label{f4}
\centerline{{\epsfxsize=9cm \epsfbox{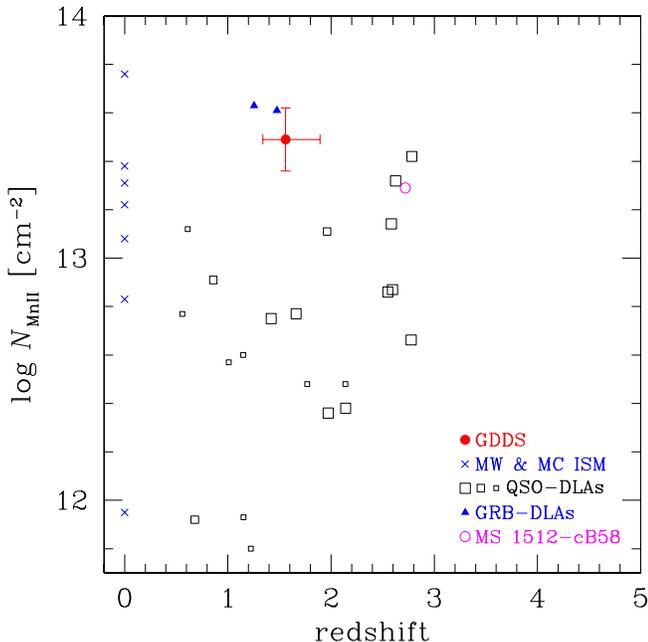}}}
\figcaption[f1]{As in Fig.~7, but for \ion{Mn}{2} column densities. The error
for \mnii~ in QSO--DLAs is on average $0.07$ dex.  Errors for local
ISM column densities, when available, are typically less than 0.1
dex.}
\end{figure}

\subsection{Metal content at high redshift}

One of the advantages of using ISM absorption lines associated with
singly ionized elements in galaxies is that no model--dependent
assumptions or calibrations, dust extinction or ionization corrections
are important when deriving the heavy element enrichment.

In Fig.~7 we show the \ion{Fe}{2} column density measured in the ISM
of different galaxies, as a function of redshift.  In our analysis we
unambiguously find that the metal enrichment of the GDDS galaxies is
high, with \feiib.  Eighteen QSO--DLAs have measured \ion{Fe}{2}
column density in the same redshift range (\zii). In this sample the
highest column density of \ion{Fe}{2} is \feii~$\simeq15.36$,
i.e. $\sim1.5$ times lower than in the GDDS \coadd.  If we extend the
range to $1.0< z<2.0$, we find 24 QSO--DLAs, and 2 systems with
comparable column density (8\% of the total, or 15\% if only $\log
N_{\rm HI}\geq 20.3$ systems are considered). This is still a
significant difference with the GDDS \coadd. On the other hand, high
values of \ion{Fe}{2} column density are found in other star--forming
galaxies: the GRB host galaxies, and the LBGs cB58 at $z=2.72$ (Table
3).

We note that when the background source is external to the galaxy (as
for QSO--DLAs) the column density of an ion is higher than when the
background source is inside the galaxy (as for the GDDS \coadd). In
the former case, the line of sight crosses the entire galaxy, whereas
in the latter only the gas in front of the stars is probed.  The
simplest situation would be that the stars are at the center,
therefore the observed column is a factor of 2 less than what a
background QSO would see. For some distribution of stars and gas, the
correction could be larger or smaller, but never downward. This would
reinforce our findings of high \ion{Fe}{2} column density in GDDS
galaxies with respect to typical \ion{Fe}{2} column densities in
QSO--DLAs.

Similarly, in Fig.~8 we show the \ion{Mn}{2} column densities in the
ISM of different galaxies. Even if the number of detections is much
lower, the trend shown by \ion{Fe}{2} is confirmed. In the whole
QSO--DLA sample, only two (out of 21, 9.5\%, or 13\% if $\log N_{\rm
HI}\geq 20.3$ systems only are considered) have \ion{Mn}{2} column
density comparable to the GDDS \coadd~(\mniib). At $1 < z < 2$, all 9
QSO--DLAs with detected \ion{Mn}{2} have $\log N_{\rm MnII}
\leq13.1$. Comparable values are found, once again, in GRB host
galaxies and cB58 (Table 3).

The \ion{Mg}{2} doublet is very strong in the GDDS \coadd~(Fig.~3),
but is highly saturated and the column density very uncertain
(\mgiib).  From this, we derive a magnesium-to-iron relative
abundance\footnote{We adopt the definition [X/Y] $=\log N_X/N_Y - \log
(X/Y)_\odot$, where $(X/Y)_\odot$ is the meteoritic solar abundance of
element $X$ with respect to element $Y$ as given by Grevesse \& Sauval
1998.} in the range $-0.84<$ [Mg/Fe] $<0.13$. In the Galactic ISM, Fe
is more depleted in dust than Mg, and [Mg/Fe] in the gas form is in the
range $0.1<$ [Mg/Fe] $<0.8$. A smaller content of Fe with respect to
Mg is also expected in the case of $\alpha-$element enhancement (Mg is an
$\alpha$ element) in metal poor systems. The low value of [Fe/Mg]
found in the GDDS \coadd~may also indicate that there is a residual
saturation in the measured \ion{Mg}{2} column density that is not
taken into account.

The \ion{Mg}{2} absorption has been extensively studied along QSO
sight lines (Rao \& Turnshek 2000; Churchill, Vogt, \& Charlton 2003;
Ding et al.~2003). These absorbing clouds have been claimed to be
associated with galaxy environment with $\log N_{\rm HI} \gsim
17.3$. When the EW of the \ion{Mg}{2} doublet is less than $\sim 1$
\AA, the absorbing gas originates in halo clouds, while for larger
EWs, the system is very likely a DLAs (Rao \& Turnshek 2000). In GRB
afterglows, the \ion{Mg}{2} absorption is, when detected, very strong
and often used to identify the redshift, since it can be detected in
the optical up to $z\sim2.2$. In Fig.~9 we show the \ion{Mg}{2}
doublet EWs seen in QSO sight--lines, GRB afterglows, and GDDS \coadd,
as a function of redshift.  The difference between QSO sight--line
studies and star--forming galaxies is apparent.

Unfortunately for lack of wavelength coverage we have no information
on the Ly$\alpha$ absorption in our GDDS spectra. Therefore the
\ion{H}{1} column density and the metal content relative to hydrogen
cannot be derived.  A rough estimate of the \ion{H}{1} column
density can be obtained using the correlation between the SFR surface
density and the \ion{H}{1} surface density, found by Kennicutt (1998)
for local disk galaxies. Our slit aperture of 0.75 arcsec corresponds
to a physical size (at $z= 1.6$) of 6.4 kpc. The signal along the $y$
direction has been extracted over 9.6 pixels (average value for the 13
galaxies) and this corresponds to 0.70 arcsec, or 5.9 kpc.  If we
assume that the detected flux comes from a $6.4\times5.9=37.7$ kpc$^2$
surface, and take the median SFR of the sample $\sim40$ M$_\odot$
yr$^{-1}$ (derived from [\ion{O}{2}] emission), we obtain a SFR
surface density:

\begin{equation}
\log \Sigma_{\rm SFR} \simeq 0.02 {\rm ~M_\odot~yr^{-1}~kpc^{-2}}
\end{equation}

\noindent
which corresponds to a \ion{H}{1} surface density of $\log \Sigma_{\rm
HI} \sim 1.6$ M$_\odot$ pc$^{-2}$ ($\pm0.3$ dex dispersion), or to a
\ion{H}{1} column density $\log N_{\rm HI} = 21.7\pm0.3$. Changing the
SFR by 20\%, would change the \ion{H}{1} of $\sim 10$\%. Taking the
\ion{Fe}{2} column density determined with the COG (\feiib), and
assuming a moderate Fe dust depletion, $\delta_{\rm Fe} \equiv \log
N_{\rm Fe,tot} - \log N_{\rm Fe,gas} =1.0$ dex (similar depletions are
found in Galactic warm disk and halo+disk gas clouds; Savage \&
Sembach 1996), we obtain a metallicity $Z/Z_\odot =
0.2^{+0.3}_{-0.1}$. The error includes 20\% and 0.2 dex dispersion in
the SFR and dust depletion, respectively, as well as the uncertainty
in the \ion{H}{1} and \ion{Fe}{2} column densities.

The median \ion{H}{1} column density for 135 QSO--DLAs with $\log
N_{\rm HI}\geq 20.3$ is $\log N_{\rm HI}\simeq 20.7$. About 1/4 of
these have $\log N_{\rm HI}\geq 21.0$ and none has $\log N_{\rm
HI}>21.85$.  For the LBG cB58 a $\log N_{\rm HI} \simeq 20.9$ have bee
reported (Savaglio et al.~2002; Pettini et al~2002). To date, this is
the only LBG for which the \ion{H}{1} column density has been
measured.  High \ion{H}{1} column densities have also been estimated
in GRB~000926 and GRB~030323 (Table 3).  In addition, for GRB~000301C
($z=2.040$) and GRB~020124 ($z=3.198$) a tentative $\log N_{\rm H I}
\approx 21.2$ and 21.7, respectively, have been reported (Jensen et
al.~2001; Hjorth et al.~2003). If we assume for the GDDS \coadd~a
\ion{H}{1} column density in the range $\log N_{\rm HI} \sim
21.0-21.7$, and apply the same dust correction as before ($\delta_{\rm
Fe}=1$ dex), we obtain $Z/Z_\odot \sim 1-0.2$.

A high metallicity would not be terribly surprising, because even if
QSO--DLAs dust--corrected metallicities are on average $\sim1/10$
solar in the redshift range $1 < z < 2$ (Savaglio 2000; Prochaska et
al.~2003b), at redshift $0.3 < z < 0.9$ higher metallicities (0.5 to 2
times solar) have been found in \ion{H}{2} regions (Kobulnicki et
al.~2003; Lilly et al.~2003). For cB58 at $z=2.72$ Pettini et
al.~(2002) report $Z/Z_\odot \sim2/5$.  Heckman et al.~(1998) have
found that in the local Universe more luminous starbursts have also
higher metallicities and are dustier. Similarly, our subsample of GDDS
galaxies is selected to be the most luminous in the sample, therefore
they could also be the most metal rich.

\begin{figure}\label{f7}
\centerline{{\epsfxsize=9cm \epsfbox{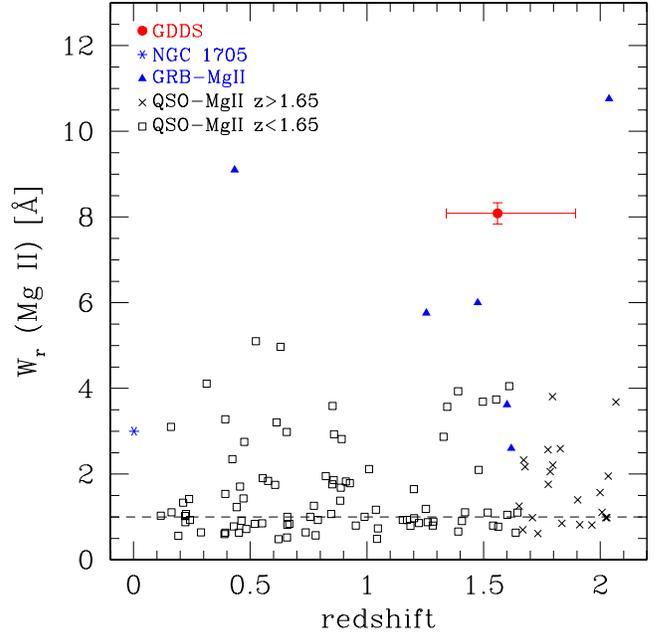}}}
\figcaption[f1]{Equivalent widths of the \ion{Mg}{2}
$\lambda\lambda2796,2803$ doublet as a function of redshift in GDDS
\coadd~galaxy, QSO--\ion{Mg}{2} absorbers, GRB afterglows and
star-forming galaxy NGC 1705. QSO--\ion{Mg}{2} absorbers at $z<1.65$
and $z>1.65$ are from Rao \& Turnshek (2000) and Steidel \& Sargent
(1992), respectively. QSO--\ion{Mg}{2} systems with total EW larger
than $W_r=1$ \AA~(dashed line) are most likely QSO--DLAs. The four GRB
detections are from, in order of increasing redshift, GRB~990712
($z=0.433$; Vreeswijk et al.~2001), GRB~020813 ($z=1.255$; Barth et
al.~2003), GRB~010222 ($z=1.475$; Jha et al.~2001; Masetti et
al.~2001; Mirabal et al.~2002; Salamanca et al.~2001), GRB~990123
($z=1.600$; Kulkarni et al. 1999), GRB~990712 ($z=1.619$; Vreeswijk et
al.~2001) and GRB~000926 ($z=2.038$; Castro et al.~2003). The EW for
NGC 1705 are from V\'azquez et al.~(in preparation).}
\end{figure}

\begin{figure}\label{f6}
\centerline{{\epsfxsize=9cm \epsfbox{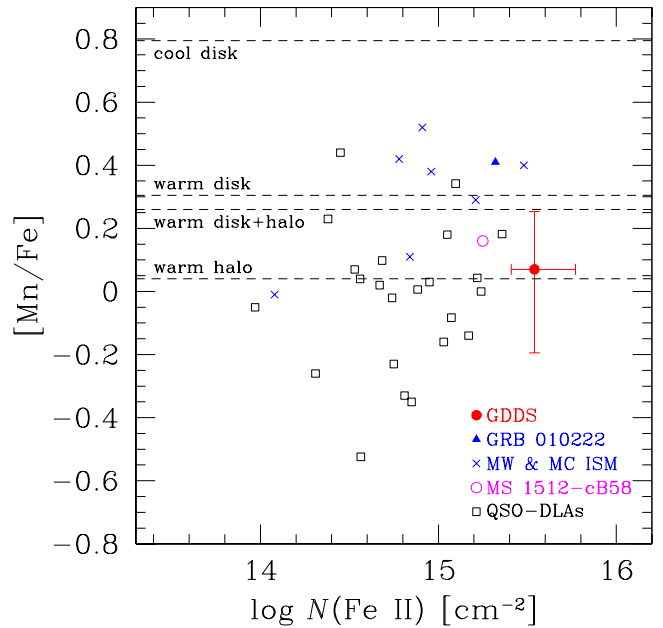}}}
\figcaption[f1]{Manganese--to--iron relative abundances as a function
of \ion{Fe}{2} column density in different high redshift galaxies and
in the MW and MCs. The dashed lines indicate [Mn/Fe] found in, from
top to bottom, cool disk, warm disk, warm disk+halo and warm halo
clouds of the MW (Savage \& Sembach 1996).}
\end{figure}

\subsection{Dust depletion and extinction in the gas}

The relative abundances of elements with different depletion levels
can be used to estimate the dust content in galaxies. For example, Zn
is typically hardly depleted in the ISM, while Fe is much more
depleted. The zinc--to--iron relative abundance is on average [Zn/Fe]
$\simeq 0.52$, in 44 QSO--DLAs ($\langle z \rangle =2.0$) with a
1$\sigma$ dispersion of 0.28 dex. This is reminiscent of the depletion
pattern in the warm halo clouds of the MW, where [Zn/Fe] $\sim 0.6$
(Savage \& Sembach 1996).

In the GDDS \coadd, the \ion{Zn}{2} doublet is off the wavelength
range, whereas \ion{Fe}{2} and \ion{Mn}{2} were both detected.
Unfortunately, both Fe and Mn are strongly affected by the presence of
dust (Savage \& Sembach 1996).  We obtain a very uncertain [Mn/Fe]
$=0.07^{+0.18}_{-0.26}$ in the GDDS \coadd~(Fig.~10). Future analysis
with more data will allow to constrain better this measurement. In 28
QSO--DLAs with $\langle z \rangle =1.59$, we found [Mn/Fe] $=-0.02$
with 0.22 dex dispersion. In the MW, the depletion patterns found by
Savage \& Sembach (1996) indicate [Mn/Fe] in the range $0.0-0.8$,
going from warm halo to cool disk clouds.  A deficiency of Mn with
respect to Fe has been found in low metallicity stars, due to
``odd--even'' effect in the iron-peak element production. This shows
[Mn/Fe] $\sim -0.4$ for [Fe/H] $<1$ and a linear increase for [Fe/H]
$>1$ up to [Mn/Fe] $\sim0.1$ for [Fe/H] $\sim 0.3$ (McWilliam, Rich, \&
Smecker--Hane 2003).

Independent of the \ion{Mn}{2} detection, the effects of dust
obscuration can be estimated by assuming that this is proportional to
the column density of \ion{Fe}{2}. We know that in the solar
neighborhood, the optical obscuration due to a gas cloud is (Bohlin,
Savage \& Drake, 1978):

\begin{equation}
A_V \simeq 0.5 \times \frac{Z}{Z_\odot}\times \frac{N_{HI}}{10^{21}} ~.
\end{equation}

\noindent
As $Z N_{\rm HI} \propto N_{\rm FeII}$, Eq.~6 can be expressed in
terms of the \ion{Fe}{2} column density in the cloud:

\begin{equation}
A_V \simeq 0.5 \times 10^{\log N_{\rm FeII}+\delta_{\rm Fe}-21-\log (\rm Fe/H)_\odot}
\end{equation}

\noindent 
where $\log (\rm Fe/H)_\odot = -4.50$ is the solar iron abundance
relative to hydrogen. For a Fe dust depletion $\delta_{\rm Fe} = 1$
dex (similar to the warm disk and warm disk+halo depletion in the MW;
Savage \& Sembach 1996), and \feiib, we obtain $A_V =
0.6^{+0.4}_{-0.2}$, regardless of the \ion{H}{1} column density. At
the wavelength of the H$\alpha$ the extinction is $\sim0.5$
magnitudes, or a factor of $\sim1.6$ in the flux. This is the value
used in \S 4 to derive SFRs from [\ion{O}{2}] emission.  For
$\delta_{\rm Fe} = 0.6$, like in the MW warm halo clouds, the optical
extinction is $A_V = 0.2^{+0.2}_{-0.1}$.

In QSO--DLAs (were iron depletion is on average $\delta_{\rm Fe} =
0.52\pm0.23$) the optical extinction is generally lower
($A_V\lsim0.1$), most likely due to an observational bias (Savaglio
2000). This is marginally supported by the work of Ellison et
al.~(2001), who tentatively found more DLAs in radio-selected QSOs
than in optically selected QSOs ($\sim2\sigma$ significant).

\subsection{SFRs from UV luminosities}

Independently of the [\ion{O}{2}] luminosity, SFRs can be derived
using the rest--frame UV luminosities. These are more affected by dust
extinction correction than those obtained using
[\ion{O}{2}]. Moreover, UV emission is less directly connected to the
instantaneous star formation, because associated with less massive
stars than the nebular emission.  However, it is interesting to
evaluate SFRs from a completely independent method. The $M_{2000}$
absolute magnitude (Table 1) gives an absolute intrinsic (dust
corrected) luminosity:

\begin{equation}
L_{2000} = 4\pi R_{10pc}^2\times 10^{-19.438-0.4(M_{2000}-A_{2000})}
\end{equation}

\noindent 
where $R_{10pc}=3.08 \times10^{19}$ cm, and $A_{2000}$ is the global
(in the gas and stars) extinction at 2000 \AA.  To derive SFRs from UV
luminosities, different conversion factors can be used, according to
different models that uses different initial mass functions (IMFs) and
metallicities. Even if the conversion factor can vary from one model
to another, Glazebrook et al.~(1999) have shown that the variation is
at most within a factor of two. For convenience, we select an
intermediate case valid for a solar metallicity and Salpeter
IMF. Glazebrook et al.~(1999) gives, for SFR = 1 M$_\odot$ yr$^{-1}$,
luminosities between 8.7 and 7.2 $\times 10^{27}$ erg s$^{-1}$
Hz$^{-1}$ for the $1500-2800$ \AA~interval.  We adopt

\begin{equation}
{\rm SFR_{2000}(M_\odot~yr^{-1})} = \frac{L_{2000}}{8\times10^{27} {\rm ~erg ~s^{-1}  ~Hz^{-1}}}~.
\end{equation}

To derive SFR$_{2000}$, we need the dust extinction at 2000
\AA~(Eq.~8).  We use two approaches: the MW extinction law and the
Calzetti law (Calzetti 2001) derived from a sample of local starburts.
The MW extinction is more appropriate for a uniform foreground screen
of dusty gas (with covering factor 1). In this case, using $A_V=0.6$
obtained from the \ion{Fe}{2} column density, we derive $E_{B-V}
=0.19$.  On the other hand, if the covering factor of the gas is $<1$,
then the dust extinction is patchy and differs on average from the one
derived from the dust depletion in the gas.  In this case we use
$E_{B-V} =0.44$, the mean value derived from a sample of local
starbursts (Calzetti et al.~1994). For both cases we obtain
$A_{2000}\sim 1.7$ magnitudes.  SFRs are given in Table 1; the mean
(and median) value is 36 M$_\odot$ yr$^{-1}$. For the subsample of
galaxies with measured [\ion{O}{2}] emission, the mean is 32 M$_\odot$
yr$^{-1}$, close to the 40 M$_\odot$ yr$^{-1}$ derived from the
[\ion{O}{2}] emission. A $\sim 1.7\sigma$ significant linear
correlation is found between the two sets of star formations (with
SFR$_{\rm [OII]} \sim 1.2$ SFR$_{2000}$).  Our SFR$_{2000}$ is
measured from the aperture corrected photometry, that includes the UV
flux from a large part of the galaxy. The consistency with values
derived from the [\ion{O}{2}] observed in the spectra (these are
associated with the smaller \ion{H}{2} regions) suggests a high degree
of mixing between the gas and the stars.

\subsection{Spectral slope in GDDS galaxies}

If the flux can be approximated by the power law $f_\lambda \propto
\lambda^\beta$, we can estimate the spectral slope $\beta$ in the
individual spectra.  The observed spectral range is not very large,
$5000-9800$ \AA, therefore the spectral slope measured from there can
be inaccurate, also because the flux calibration in MOS data is not
perfect. However, we can use the observed Vega $V$ and $I$ magnitudes
to derive $\beta$. If the AB magnitudes are $V_{AB} = V+0.02$ and
$I_{AB} = I+0.45$, then $\beta$ is:

\begin{equation}
\beta = 0.4 \times \frac{I-V+0.43}{\log (\lambda_V/\lambda_I)} -2
\end{equation}

\noindent
where $\lambda_V=5500$ \AA~and $\lambda_I=8000$ \AA. For a galaxy at
$z=1.6$, $\beta$ would measure the spectral slope in the rest--frame
interval $2100-3100$ \AA.  For the individual 13 GDDS galaxies we
obtain a mean value $\langle \beta \rangle =-1.15$ with a large
dispersion of 0.9. In local star--forming galaxies, Calzetti et
al. (1994) have measured $\beta$ in a sample of 39 local starbursts in
the wavelength range $1250-2600$ \AA. From this sample we derive a
mean value $\langle \beta \rangle = - 1.05$, with a dispersion of
0.65.  At high redshift Shapley et al.~(2003) derived $\beta$ in LBGs
using the intrinsic UV colors ($\lambda \sim1200-1800$ \AA) and found
$-1.1 < \beta < -0.7$.  To compare our spectral slope to those found
at lower wavelengths, we computed spectra using stellar synthesis
models. We assumed a simple stellar population, with Salpeter IMF and
continuous star formation. The spectral slope in the range
$\lambda\sim1200-2000$ \AA~is slightly flatter than in the range
$\lambda\sim2000-3000$ \AA, due the contamination by strong stellar
absorptions at $\lambda=1200-2000$ \AA.  

By applying the MW and starburst dust extinction laws, as described in
the previous section, we calculated an intrinsic spectral slope in the
interval $\lambda\sim2100-3100$ \AA~for our GDDS galaxies $\beta_i =
-2.1$ and $-2.0$, respectively.

Dunne, Eales, \& Edmunds (2003) have found from submillimetre surveys
that dust content was much higher in the past, therefore most of the
metal rich systems cannot be found in surveys that are too shallow, or
too sensitive to dust obscuration. GDDS cannot detect galaxies or
regions of galaxies that are strongly obscured, so it is certainly
affected by dust, but probably much less than other galaxies surveys,
because it is unbiased and highly complete for galaxies with $K<20.6$.

\section{Summary and conclusions}

The GDDS (Abraham et al.~in preparation) is a public survey that uses
GMOS at Gemini, combined with the Nod \& Shuffle technique, to study
the properties of galaxies in the redshift desert ($0.8<z<2.0$). The
GDDS observed galaxies with $I_{AB} \leq 24.8$ and $K<20.6$, and is
the most sensitive survey focused on the redshift desert obtained so
far. The $K20$ survey obtained at the Very Large Telescope (VLT) is
aiming at a similar scientific objective (Cimatti et al. 2002a) with
82 spectroscopically confirmed galaxies in the redshift desert in 52
arcmin$^2$.  In the first two GDDS fields observed, 60.5 arcmin$^2$
and 59.5 hours exposure time, we identified 68 galaxies with
spectroscopically--confirmed redshift. Among these, 34 are at
$1.3<z<2.5$ (0.56 galaxies arcmin$^{-2}$), whereas the $K20$ (17 VLT
nights) identified 17 galaxies (0.32 galaxies arcmin$^{-2}$) in the
same redshift interval, showing the high efficiency of GDDS.

In this work, we have selected the UV brightest (AB absolute magnitude
$M_{2000}<-19.65$) subsample with 13 galaxies that are in the redshift
range \zii~($\langle z \rangle=$ \zav) in order to study heavy
elements in the neutral ISM. The selected galaxies are the most
massive in the sample ($M\gsim 10^{10}$ M$_\odot$) and represents
between 25 and 38\% of the total number of galaxies found in the
$z>1.3$ range (depending whether galaxies with tentative or
photometric--only redshifts are included or not). Only three of these
13 galaxies have $K<20.0$.

We detected strong [\ion{O}{2}] emission associated with \ion{H}{2}
regions, with absolute luminosities ranging $5-40\times 10^{41}$ erg
s$^{-1}$, and roughly estimated SFR $=13-106$ M$_\odot$ yr$^{-1}$ and
$<$SFR$>\sim40$ M$_\odot$ yr$^{-1}$. The mean value is close to the
mean value derived using UV luminosities found from the
aperture--corrected photometry ($32$ M$_\odot$ yr$^{-1}$). This
supports the assumption that on average the gas in the \ion{H}{2}
regions and the UV emitting stars are well mixed.

The stack of the 13 spectra shows \ion{Fe}{2}, \ion{Mg}{2},
\ion{Mn}{2}, and \ion{Mg}{1} absorption.  The column density of
\ion{Fe}{2} and \ion{Mn}{2} provide a good measure of the total column
density of iron and manganese in the neutral ISM.  We found \feiib~and
\mniib. These values are larger than what typically measured in
absorption systems along QSO sight lines (QSO--DLAs). Only 2 out of 24
at $1.0< z <2.0$ have similar \ion{Fe}{2} column densities.  On the
other hand, high column densities are found in other star--forming
objects. The Lyman break galaxy cB58 has \feii~$=15.25$, while GRB
host galaxies (GRB--DLAs) have \feii~$=14.8-15.9$. From the depth of the
\ion{Mg}{2} absorption, we derived a conservative lower limit to the
ISM filling factor of 70\%.

The high filling factor of the ISM and the consistency between SFRs
derived from [\ion{O}{2}] and UV emission, suggests a homogeneous
distribution of stars and gas. If we assume that the mean \ion{H}{1}
column density in the ISM is $\log N_{\rm HI} <21.7$ (as in 98\% of
QSO--DLAs), the mean metallicty would be $Z/Z_\odot>0.2$ for a
moderate Fe dust depletion correction ($\delta_{\rm Fe} = 1$ dex).
High metallicities ($Z/Z_\odot =0.5-2$) are also found in \ion{H}{2}
regions of galaxies at $0.3 < z < 0.9$ (Kobulnicky et al. 2003; Lilly
et al.~2003).

From the \ion{Fe}{2} column density, we estimated an optical dust
obscuration $A_V \sim 0.6$ mag. The mean spectral slope in our GDDS
galaxies around the $2100-3100$ \AA~ interval is $\beta\sim-1.15$,
from which we derive an intrinsic dust corrected mean spectral slope
$\beta \sim -2$.

When completed, the GDDS will provide observations of two additional
fields.  From the full data set we will detect or give significant
limits on other absorption lines. For instance, the \ion{Zn}{2}
detection at $\sim 2000$ \AA, or \ion{Si}{2} at $\lambda\sim 1800$
\AA, are powerful diagnostics for metal enrichment and dust depletion.
GDDS galaxies and GRB--DLAs are clearly probing high$-z$ galaxies with
a strong star formation activity, for which metal enrichment and/or
dust obscuration can be higher than in the QSO--DLA population. Our
conclusion is that direct detections of galaxies, and indirect through
QSO absorption lines, provide a more complete picture of
the galaxy census in the high redshift Universe. The former can be
seen only if star formation is high, the latter only when metal
enrichment is low.

\acknowledgments 

This paper is based on observations obtained at the Gemini
Observatory, which is operated by the Association of Universities for
Research in Astronomy, Inc., under a cooperative agreement with the
NSF on behalf of the Gemini partnership: the National Science
Foundation (United States), the Particle Physics and Astronomy
Research Council (United Kingdom), the National Research Council
(Canada), CONICYT (Chile), the Australian Research Council
(Australia), CNPq (Brazil), and CONICET (Argentina)

We are grateful to the entire staff of the Gemini Observatory for the
kind hospitality and support during our visits. We thank Matt
Mountain, Jean--Ren\'e Roy, and Doug Simons for support of the program
(including the Director's Discretionary Time allocated to GDDS),
Matthieu Bec and Tatiana Paz for technical modifications to the GMOS
telescope control system, and Richard Wolff for work on the detector
controller.  We acknowledge the anonymous referee for thoughtful
comments that greatly improved the discussion.  We thank A. Aloisi,
C. Hoopes, N. Masetti, and N. Walborn for interesting
insights. S.S. acknowledges generous funding from the David and
Lucille Packard Foundation.  H.--W.C. acknowledges support by NASA
through a Hubble Fellowship grant HF-01147.01A from the Space
Telescope Science Institute, which is operated by the Association of
Universities for Research in Astronomy, Incorporated, under NASA
contract NAS5-26555.

{}

\end{document}